\documentclass[twocolumn,prl,amsmath,amssymb,showpacs]{revtex4-1}
\usepackage{epsf}      
\usepackage{graphicx}
\usepackage{color}
\usepackage{soul}
\usepackage{gensymb}
\usepackage{sidecap}
\usepackage{amsmath}
\usepackage{multirow}

\begin{document}

\title{Unraveling the magnetic interactions and spin state in insulating Sr$_{2-x}$La$_x$CoNbO$_6$}
\author{Ajay Kumar}
\affiliation{Department of Physics, Indian Institute of Technology Delhi, Hauz Khas, New Delhi-110016, India}
\author{R. S. Dhaka}
\email{rsdhaka@physics.iitd.ac.in}
\affiliation{Department of Physics, Indian Institute of Technology Delhi, Hauz Khas, New Delhi-110016, India}
\date{\today}      

\begin{abstract}

We investigate the structural, magnetic and spin-state transitions, and magnetocaloric properties of Sr$_{2-x}$La$_x$CoNbO$_6$ ($x=$ 0--1) double perovskites. The structural transition from tetragonal to monoclinic phase at $x$ $\geqslant$ 0.6, and an evolution of (101)/(103) superlattice reflections and Raman active modes indicate the enhancement in the B-site ordering with $x$. The magnetic susceptibility data reveal the transition from weak ferromagnetic (FM) to antiferromagnetic (AFM) ordering for $x$ $\geqslant$ 0.6 with T$_{\rm N}$$\approx$9--15~K. Interestingly, the La substitution drives towards more insulating state due to increase in high-spin Co$^{2+}$, whereas a spin-state crossover is observed in Co$^{3+}$ from high-spin to intermediate/low-spin states with $x$. We discuss the correlation between complex magnetic interactions and the presence of various Co spin-states in the system. Moreover, the emergence of metamagnetic nature due to the competition between FM and AFM interactions as well as crossover from conventional to inverse magnetocaloric effect have been demonstrated by detailed analysis of temperature and field dependent change in magnetic entropy.  

\end{abstract}

\maketitle

\section{\noindent ~Introduction}

The spin-state crossover at 80--90~K and its crucial role in understanding the unusual magnetism and electronic transport in LaCoO$_3$ are most famous long-standing problems in solid state physics \cite{Raccah_PR_67, Bhide_PRB_72, Asai_PRB_89}. The presence of various spin-states of Co$^{3+}$ ion i.e., low spin (LS; t$_{2g}^6$ e$_g^0$), intermediate spin (IS; t$_{2g}^5$ e$_g^1$), high-spin (HS; t$_{2g}^4$ e$_g^2$), and/or mixed states is still under debate in spite of the extensive experimental and theoretical studies \cite{Raccah_PR_67, Bhide_PRB_72, Asai_PRB_89, Zhuang_PRB_98, Yan_PRB_04, 
 Korotin_PRB_96, Saitoh_PRB_97, Chainani_PRB_92}. A strong competition between crystal field splitting ($\Delta_{\rm cf}$) and Hund's exchange energy ($\Delta_{\rm ex}$) results in the considerably small energy difference between HS and LS states \cite{Raccah_PR_67, Bhide_PRB_72, Asai_PRB_89}. Also, the IS state found to be energetically close to the LS state when the hybridization between Co e$_g$ and O 2$p$ orbitals has been considered \cite{Korotin_PRB_96}. In order to understand the temperature dependent magnetic susceptibility ($\chi$-T) behavior of LaCoO$_3$, it has been suggested that the presence of HS states favors the antiferromagnetic (AFM) exchange interactions, whereas  the IS states of Co$^{3+}$ show the ferromagnetic (FM) interactions \cite{Saitoh_PRB_97, Troyanchuk_PSS_05, Yoshii_JAC_2000}. Hence, the presence of mixed states results in the competition between FM and AFM interactions, which depends on their relative population. In this direction, double perovskite oxides with general formula A$_2$BB$^\prime$O$_6$ (A: rare earth/alkali earth metals, B/B$^\prime$: transition metals) have attracted great attention due to an extra degree of freedom of tuning the B-site ordering \cite{Galasso_JPC_62, Anderson_SSC_93, King_JMC_10, Vasala_SSC_15}. Here the rock salt like B-site ordering is governed by two major factors: difference in the valence state of  two B-site cations ($\Delta$V$_{\rm B}$) and their ionic mismatch ($\Delta$r$_{\rm B}$). It is important to note that the large $\Delta$V$_{\rm B}$ and $\Delta$r$_{\rm B}$ favor the ordering in the crystals due to reduction in the coulombic repulsion energy and lattice strain, respectively \cite{Anderson_SSC_93, King_JMC_10, Vasala_SSC_15}. The degree of B-site ordering in these materials governs most of their magnetic, transport, and electronic properties \cite{Jung_PRB_07, Bos_PRB1_04, Narayanan_PRB_10, Sarma_PRL_07, Meneghini_PRL_09} and consequently their technological aspects, namely in photocatalysis, photovoltaics, solid oxide fuel cells, etc. \cite{Yin_EES_19, Kangsabanik_PRM_18, Huang_CM_09, Yoo_RSC_14, Kobayashi_nature_98, Sun_AM_18}. The extent of ordering can be quantified by long range order parameter, S = 2$z-$ 1 \cite{Vasala_SSC_15, Meneghini_PRL_09}, where $z$ is the fractional occupancy of B/B$^{\prime}$ cations at their given Wyckoff positions. The values of S = 1 and 0 indicate completely ordered and disordered configurations, respectively.

Interestingly, due to the combined effect of the multiple spin-states of Co and flexibility in the B-site cationic ordering, Co-based double perovskite oxides  are particularly important to understand their unusual physical properties with aliovalent substitution at A-site \cite{Bos_PRB1_04, Ding_PRB_19, Bos_PRB_04, Narayanan_PRB_10, Haripriya_PRB_19}. However, the substitution at A- and/or B- site(s) is limited by the ionic mismatch between the different atoms, which can be estimated by the Goldschmidt tolerance factor, $\tau$, where the deviation of $\tau$ from the unity gives an estimation of the chemical pressure in the system and hence possibility of  lowering in the crystal symmetry. Moreover, the Co$^{2+}$ is energetically favorable to exist in the HS state due to weaker crystal field effect as compared to Co$^{3+}$ \cite{Mabbs_Dover_08}. Also, due to the presence of triply degenerate $^4T_1$ ground term, HS Co$^{2+}$ systems in octahedral coordination results in the unquenched orbital magnetic moment, which make this system more interesting \cite {Lloret_ICA_08, Viola_CM_03}. Further, octahedrally coordinated HS Co$^{2+}$ shows the strong AFM exchange interactions and thus a mixture of Co$^{3+}$ and Co$^{2+}$ lead to the several competitive interactions, which give rise to the exotic physical properties \cite{Yuste_DT_11, Bos_CM_04}. For example, a significantly different behavior of electronic transport has been reported in La$_{1+x}A_{1-x}$CoRuO$_6$ ($A=$ Ca, Sr) with hole and electron doping \cite{Bos_PRB1_04, Bos_CM_04}. More recently, the spin-orbit coupling in Ir and magnetic interactions between Co 3$d$ and Ir 5$d$ (spatially extended orbitals) moments have been studied along with spin-state transition between HS/LS-Co$^{2+}$/LS-Ir$^{4+}$ to HS-Co$^{3+}$/LS-Ir$^{5+}$ in La$_{2-x}$Sr$_x$CoIrO$_6$ systems \cite{Lee_PRB_18, Marco_PRB_15, Kolchinskaya_PRB_12, Vogl_PRB_18, Coutrim_PRB_16}. Also, the effect of small ionic mismatch between B-site cations and mixed valance states in various Co-based double pervoskites $A_2$Co$B$O$_6$ ($A=$ Sr, Ca, Ba, Y, La and $B=$ Mn, Fe) results in exciting phenomenon like large magnetoresistance, exchange bias, spin-glass, cluster-glass, memory effect, multiferroicity, magnetocaloric, and thermoelectric effects \cite{Madhogaria_PRB_19, SahooPRB_19, Pradheesh_EPJB_12, Coutrim_PRB_19, MurthyJPD14, SharmaAPL13, TanwarPRB19}, which are crucial for various device applications. The observation of half-metallicity in Sr$_2$FeMoO$_6$ with high ferromagnetic Curie temperature has also opened the possibility of room temperature spintronics \cite{Kobayashi_nature_98, ErtenPRL11}, where the knowledge of degree of disorder is important for high spin polarization. However, the presence of magnetic atoms at both B and B$^\prime$ sites (in some cases A-site as well \cite{Ding_PRB_19}) and complex magnetic interactions between them make it difficult to study the role of individual atom in controlling their physical properties.

Therefore, in order to understand the magnetic/spin-states of Co ion in the octahedral environment and its evolution with temperature, and B/B$^\prime$ ordering, it is vital to investigate the Co-based double perovskites with other non-magnetic atoms; for example Ti$^{4+}$ (3$p^6$) \cite{Yuste_DT_11,Shafeie_JSSC_11}, Nb$^{5+}$ (4$d^0$) \cite{Bos_PRB1_04, Kobayashi_JPSJ_12, Yoshii_JSSC_2000}, which are largely unexplored. A large discrepancy in the effective magnetic moment ($\mu_{\rm eff}$) and hence the spin-states of Co exist in literature. Recently, we have reported that Nb substitution at Co site in LaCoO$_3$ convert Co$^{3+}$ to Co$^{2+}$ and stabilize them in HS state along with structural transition \cite{ShuklaPRB18}. Whereas, the substitution of Sr/Nb in LaCoO$_3$ stabilize Co$^{3+}$ in a mixed state of IS and HS \cite{RaviJALCOM18, ShuklaJPCC19}. Moreover, in double pervoskites the $\mu_{\rm eff}$ of Sr$_2$CoNbO$_6$ found to be equal to that of spin only moment 4.9 $\mu_B$, which confirms the presence of Co$^{3+}$ only in HS states \cite{Yoshii_JAC_2000}. While the $\mu_{\rm eff}$ is found to be 2.06 $\mu_B$ and 1.91 $\mu_B$ in refs.~\cite{Azcondo_Dalton_15} and \cite{Kobayashi_JPSJ_12}, respectively, which suggest the presence of LS--HS or LS--IS states. Also, the Co$^{3+}$ in mixed HS and IS states has been reported in Ba$_2$CoNbO$_6$ \cite {Yoshii_JSSC_2000}. Interestingly, the bifurcation in ZFC--FC curves of Ba$_2$CoNbO$_6$ suppresses with applied magnetic field at around H$>$10~kOe \cite{Yoshii_JSSC_2000}; while no such behavior was observed in the $\chi-$T curves of Sr$_2$CoNbO$_6$ \cite{ Yoshii_JAC_2000}. Also, the (La$A$)CoNbO$_6$ ($A=$ Ca, Sr, and Ba) show the magnetic frustration which increases with the size of A-site cations and a significant orbital magnetic moment has been reported due to HS Co$^{2+}$ in the octahedral environment \cite{Bos_PRB1_04}. However, aliovalent substitution at A site is largely unexplored in Sr$_2$CoNbO$_6$, where the substitution of La$^{3+}$ (5$p^6$) at Sr$^{2+}$ (4$p^6$) site (both non-magnetic) can change Co$^{3+}$ to Co$^{2+}$ i.e., $\Delta$V from 2 to 3, as the Nb$^{5+}$ valence state remains unaltered, and hence control the B-site ordering due to the moderate ionic mismatch between Co and Nb. Moreover, the change in the magnetic entropy ($\Delta$S) derived from the magnetization measured at various temperatures across the transition is more sensitive to the complex magnetic behavior \cite{Phan_JMMM_07}.  

In this paper we investigate the structural, magnetic, magnetocaloric and electronic transport behavior in Sr$_{2-x}$La$_x$CoNbO$_6$ ($x=$ 0--1) double perovskites. We found a structural phase transition, and the evolution of (101)/(103) superlattice reflections and appearance of Raman active modes indicate a systematic enhancement in the B-site ordering with $x$. A long range antiferromagnetic ordering is clearly evident in $\chi-$T data for $x$ $\geqslant$ 0.6 samples with the T$_{\rm N}$ value of 9--15~K. Our isothermal magnetization study reveals the low temperature weak FM interactions for $x$ $\leqslant$ 0.4 and a weak metamagnetic signature for $x$ $\geqslant$ 0.6. The substitution of La at Sr site changes Co$^{3+}$ (for $x=$ 0) to Co$^{2+}$ (for $x=$ 1). The orbital contribution in magnetic moment in case of Co$^{2+}$ and spin-state of Co$^{3+}$ are estimated from the $\chi-$T measurements. The competition between FM and AFM interactions has been found by analyzing the temperature and field dependent $\Delta$S. Moreover, a drastic reduction in the electronic conductivity with increase in $x$, due to the conversion of Co$^{3+}$ into Co$^{2+}$ has been discussed.

\section{\noindent ~Experimental}

 Polycrystalline Sr$_{2-x}$La$_{x}$CoNbO$_{6}$ ($x=$ 0--1) samples were synthesized by solid-state route. We used all the precursors as purchased except Lanthanum (III) Oxide (La$_{2}$O$_{3}$), which was pre-heated at 900$\degree$C for 3~hrs to ensure the dehydration \cite{Neumann_TA_06}. The stoichiometric amount of strontium carbonate (SrCO$_{3}$), niobium (V) oxide (Nb$_{2}$O$_{5}$), cobalt(II,III) oxide (Co$_{3}$O$_{4}$) and dried La$_{2}$O$_{3}$ (all were from Sigma/Alfa with purity $\ge$ 99.99\%) were ground in a mortar-pastel for around 5~hrs for the homogeneous mixing. Then, pressed the resultant powder into pellets with hydraulic press at 2500 psi and calcined at 900$\degree$C for 2~hrs and finally sintered at 1300$\degree$C for 48~hrs with intermediate grindings \cite{Bos_PRB_04}. The structural investigation has been done by x-ray diffraction (XRD) measurements at room temperature using Panalytical Aeris diffractometer in Bragg- Brentano geometry with Cu-K$_{\rm \alpha}$ ($\lambda$=1.5406 \AA) radiation and 40~keV accelerating voltage. For the Rietveld refinement of XRD patterns of all the samples, we used the pseudo voigt peak shape and linear interpolation between the set background points in  FullProf suite \cite{Carvajal_PB_93}. We use a Renishaw inVia confocal Raman microscope to perform the Raman spectroscopy measurements in backscattering geometry with a 532~nm excitation wavelength, 2400 lines per mm grating and 1~mW laser power. The temperature and field dependent magnetization (M--T and M--H) measurements were carried out using superconducting quantum interference device (SQUID) from Quantum Design, USA. The virgin magnetization isotherms at different temperatures and temperature dependent resistivity ($\rho$--T) data were collected using physical property measurement system (PPMS) from Quantum Design, USA.

\section{\noindent ~Results and discussion}

\begin{figure}[h]  
\includegraphics[width=3.4in]{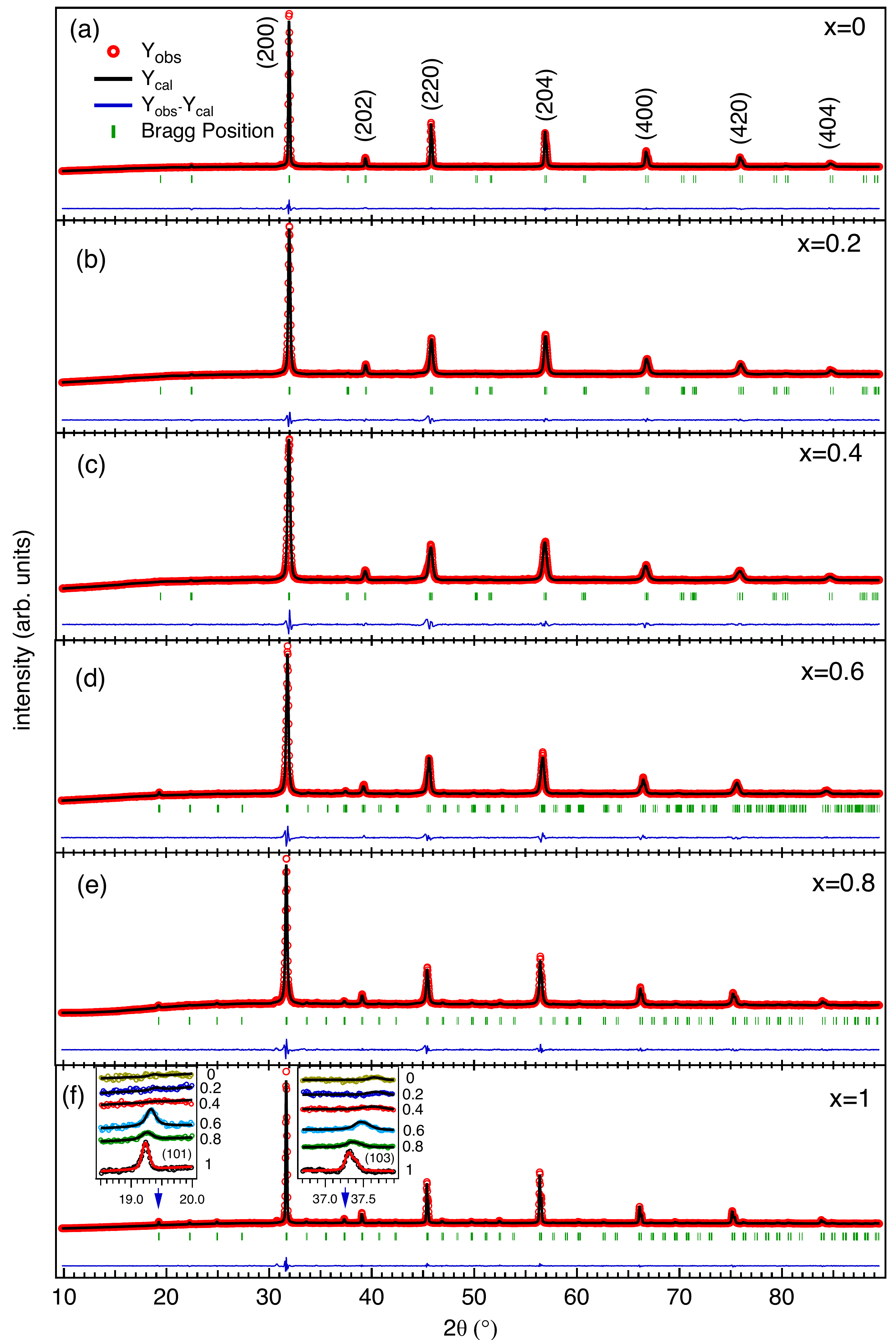}
\caption {(a--f) The Rietveld refinement of XRD data of Sr$_{2-x}$La$_x$CoNbO$_6$ ($x=$ 0--1), where the observed data points, simulated curve, Bragg positions and difference in observed and simulated data are shown by open red circles, continuous black line, vertical green bars and continuous blue line, respectively. Insets of panel (f) show the evolution of the superlattice reflections (101) and (103) with La concentration.} 
\label{fig:XRD-all}
\end{figure}

In Figs.~\ref{fig:XRD-all}(a--f) we present the Rietveld refinement of room temperature XRD patterns of Sr$_{2-x}$La$_x$CoNbO$_6$ ($x=$ 0--1) samples, which show the tetragonal structure ($I4/m$) up to $x \leqslant$ 0.4 having a$^0$a$^0$c$^-$ tilt system. While we found a monoclinic ($P2_1/n$) distortion in the crystal structure for $x\geqslant$ 0.6 samples having b$^-$b$^-$c$^-$ tilt \cite{Bos_PRB_04} along with an increase in lattice parameters. The octahedral tilting is described by the Glazer notations \cite{Glazer_AC_72, Woodward_AC_97}, where letter denotes the magnitude of rotation of octahedron about an axis relative to its magnitude of rotation about another crystallographic axis, and its superscript (+/-)  indicates whether the rotation of the adjacent octahedron is in the same direction (+) or opposite (-). The absence of any rotation of octahedron about a particular axis is denoted by 0 superscript. The difference in the effective ionic radii of La$^{3+}$(1.36~\AA; 12 coordinated), Sr$^{2+}$(1.44~\AA; 12 coordinated), Co$^{3+}$(0.545~\AA~ for LS and 0.610~\AA~ for HS; 6 coordinated) and Co$^{2+}$(0.650~\AA~ for LS and 0.745~\AA~ for HS; 6 coordinated) \cite{Shannon_AC_76} is the key factor for this lowering in the crystal symmetry and the unit cell expansion with increase in the La concentration. As stated above, substitution of smaller La$^{3+}$ cations at larger Sr$^{2+}$ site enhance the concentration of Co$^{2+}$  ions in the same proportion, which are larger than Co$^{3+}$ ions. Thus, the cumulative effect of both Sr$^{2+}$$\rightarrow$ La$^{3+}$ and Co$^{3+}$$\rightarrow$Co$^{2+}$ governs the change in the crystal symmetry of these materials. This can be understood by the tolerance factor $\tau$ values, as listed in Table~I for all the samples along with other Rietveld refined parameters, keeping in mind the low sensitivity of the conventional XRD for oxygen and hence fallacy in the bond lengths and bond angles containing the oxygen atoms. We calculate the $\tau$ by taking the average ionic radii of LS and HS states for Co$^{3+}$ and HS state for Co$^{2+}$. The decrease in $\tau$ value with $x$ below unity clearly indicates the strong possibility of lowering in the crystal symmetry. Interestingly, the evolution of the superlattice reflections (101) and (103), originating from the alternating CoO$_6$ and NbO$_6$ octahedra, suggests the enhancement in the B-site ordering, as shown in the insets of Fig.~\ref{fig:XRD-all}(f). This is consistent as higher concentration of Co$^{2+}$ ions due to substitution of Sr$^{2+}$ ions by La$^{3+}$ ions results in the increment in $\triangle$V, which is the key factor to favor the ordering at the B-site. Here, the percentage of disorder can be quantitatively extracted by refining the occupancy of Nb and Co atoms at their given Wyckoff positions, see S parameter in Table~I. This indicate the transition from completely disordered ($x=$ 0) to almost ordered ($x=$ 1) structure with La substitution \cite{Bos_PRB1_04, Yoshii_JAC_2000}. We note here that the S parameter increases systematically with $x$, whereas the intensity of the ordered peak for the $x=$ 0.8 sample is less than that of the $x=$ 0.6 sample, see insets of Fig.~\ref{fig:XRD-all}(f). This discrepancy is possibly due to different degree of octahedral tilting, which can also lead to the evolution of these reflections \cite{Barnes_AC_06}.

\begin{table*}
		\label{tab_rietveld}
		\caption{Rietveld refined order parameter (S), lattice parameters, calculated tolerance factor, selected bond lengths (\AA) and bond angles (degree) along with fitting reliability parameters of Sr$_{2-x}$La$_x$CoNbO$_6$ ($x=$ 0--1) samples.}
		
		\begin{tabular}{p{3cm}p{2cm}p{2cm}p{2cm}p{2cm}p{2cm}p{2cm}}
\hline
		\hline
	$x$ &0&0.2&0.4&0.6&0.8&1\\
\hline

$\tau$ &0.9997 &0.9886&0.9778&0.9671&0.9565&0.9462 \\
	Phase & Tetragonal & Tetragonal &Tetragonal &Monoclinic & Monoclinic & Monoclinic \\
		
S&0.00 &0.02&0.14& 0.62 &0.80&0.94\\

		a(\AA) &5.602 & 5.595&5.600&5.622&5.641&5.648\\
	
		b(\AA) &5.602 & 5.595 &5.600 &5.618 &5.642 &5.653\\
		
                 c(\AA) &7.921&7.932&7.948 &7.977&7.984&7.984\\

$\beta$($^0$) & 90 &90  &90 &89.7 &89.9 &89.9\\

Volume (\AA$^3$) &248.59& 248.32&249.23&251.92&254.09&254.92\\

\textless Co---O\textgreater$_6$ &1.98&1.99&1.99& 1.99 & 1.92 & 2.00\\
\textless Nb---O\textgreater$_6$&1.99&1.99&2.02&2.07& 2.11& 2.07\\
\textless La/Sr---O\textgreater$_{12}$&2.80&2.81&2.81&2.83&2.84&2.84\\
\textless Co---Nb\textgreater$_{6}$& 3.961 & 3.960 & 3.964 & 3.979 & 3.990 & 3.994\\
\textless Co---O---Nb\textgreater$_6$&178.7&169.1&166.3&160.7&161.1&158.4\\
$\chi^2$ &1.83 & 2.04 & 2.39 & 2.27 &1.53 &1.79\\
R$_p$ &1.92\% & 2.06\% & 2.31\% & 2.40\% &3.14\% &2.48\%\\
R$_{wp}$ &2.65\% & 2.93\% & 3.33\% & 3.44\% & 4.18\% &3.44\%\\
R$_{F}$ &4.70\% & 2.57\% & 2.06\% & 4.69\% &5.13\% &4.81\%\\
\hline
\hline
\end{tabular}
\end{table*}

\begin{figure}[h]
\includegraphics[width=3.35in]{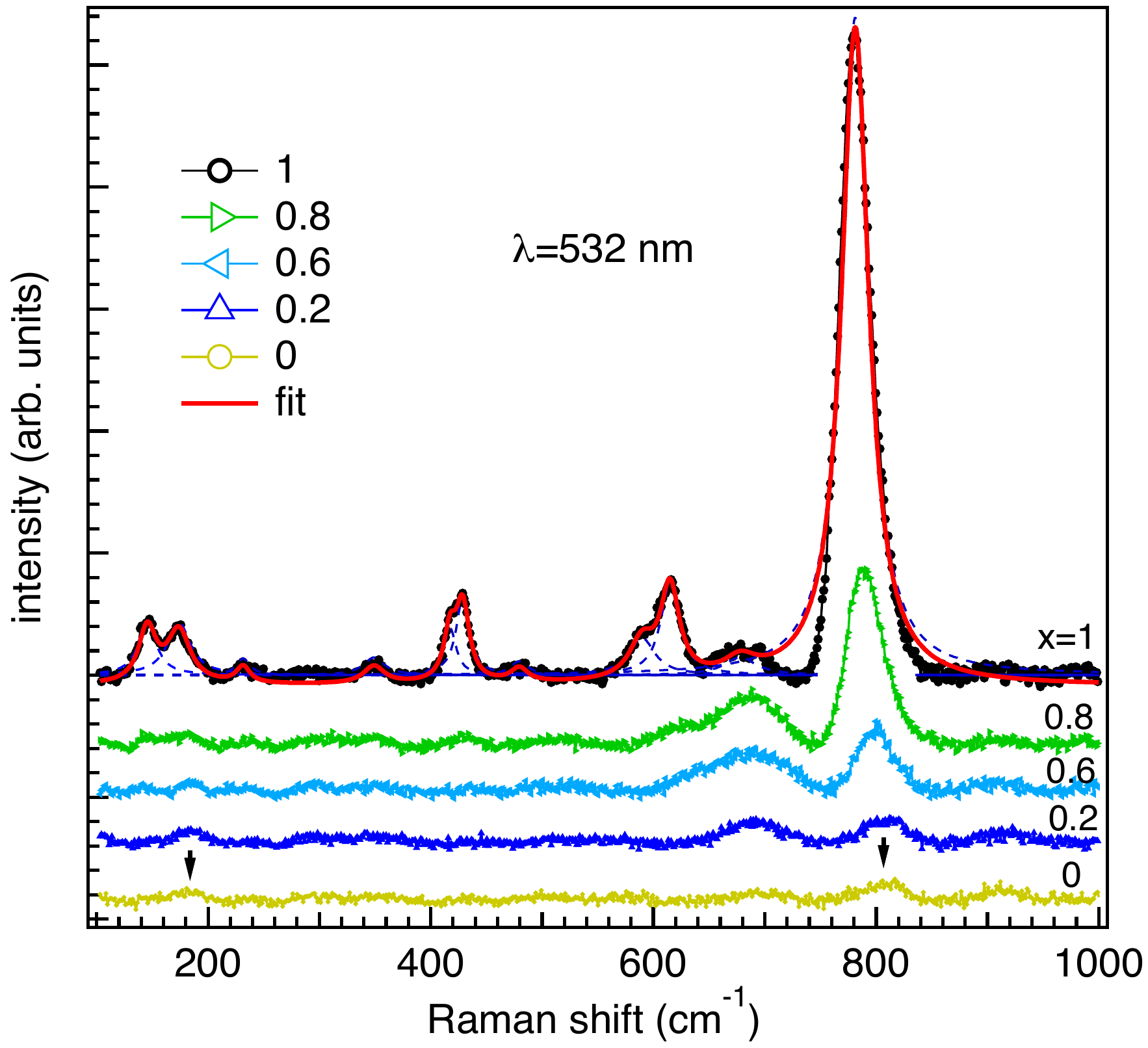}
\caption{Raman spectra of Sr$_{2-x}$La$_x$CoNbO$_6$ ($x=$ 0--1) measured at room temperature using 532 nm excitation wavelength. Red solid line in case of $x=$ 1 sample represents the Lorentzian fitting of the data points. The curves have been vertically shifted for the clear presentation.}
\label{Fig9_Raman}
\end{figure}

To further check the microstructure i.e., tilting, rotation and/or ordering of (B/B$^\prime$)O$_6$ octahedra, we show Raman spectra in Fig.~\ref{Fig9_Raman} where a sharp enhancement as well as evolution of new Raman active modes can be clearly observed with $x$. The group theory predict nine Raman active modes for $I4/m$ (tetragonal) crystal structure as $\Gamma_g$($I4/m$) = $\nu_1$(A$_g$)+ $\nu_2$(A$_g$+ B$_g$)+ $\nu_5$(B$_g$+E$_g$)+ T(B$_g$+E$_g$)+ L(A$_g$+E$_g$), where  $\nu_1$, $\nu_2$ and $\nu_5$ are oxygen symmetric stretching, asymmetric stretching and bending modes, respectively, i.e., internal modes of octahedron, and T and L are translational and vibrational modes due to the motion of A-site cations and rotation of octahedra, respectively, i.e., external modes \cite{Andrews_Dalton_15}. For the $x=$ 0 sample, we observe two weak Raman modes at around 190 and 800 cm$^{-1}$ (as indicated by arrows), which can be ascribed to the T and $\nu_1$ modes, respectively. This discard the possibility of centrosymmetric (Raman inactive) Pm$\bar{3}$m (cubic) structure for the $x=$ 0 sample, as claimed in refs. \cite{Yoshii_JAC_2000, Azcondo_Dalton_15}. Notably for the ordered $P2_1/n$ (monoclinic) structure, a total of 24 Raman active modes are predicted having irreducible representation as $\Gamma_g$ ($P2_1/n$) = $\nu_1$(A$_{g}$ + B$_g$) + 2$\nu_2$(A$_{g}$ + B$_g$) + 3$\nu_5$(A$_g$ + B$_{g}$) + 3T(A$_g$ + B$_{2g}$) + 3L(A$_g$ + E$_g$) \cite{Andrews_Dalton_15, Iliev_PRB_07}. We observe a most intense band around 800 cm$^{-1}$ due to oxygen symmetric stretching ($\nu_1$), whereas modes between 550--650 and 330--500 cm$^{-1}$ represent the oxygen asymmetric stretching ($\nu_2$) and bending ($\nu_5$) modes, respectively, along with translational (T) modes in the range between 100--250 cm$^{-1}$ \cite{Andrews_Dalton_15}. The presence of several Raman active modes in case of the $x=$ 1 sample directly confirms the fact of enhancement in Co/Nb ordering and consequently lowering in crystal symmetry with La substitution \cite{Prosandeev_PRB_05}, as also evident from the XRD  analysis. 

 \begin{figure}
\centering
\includegraphics[width=3.35in]{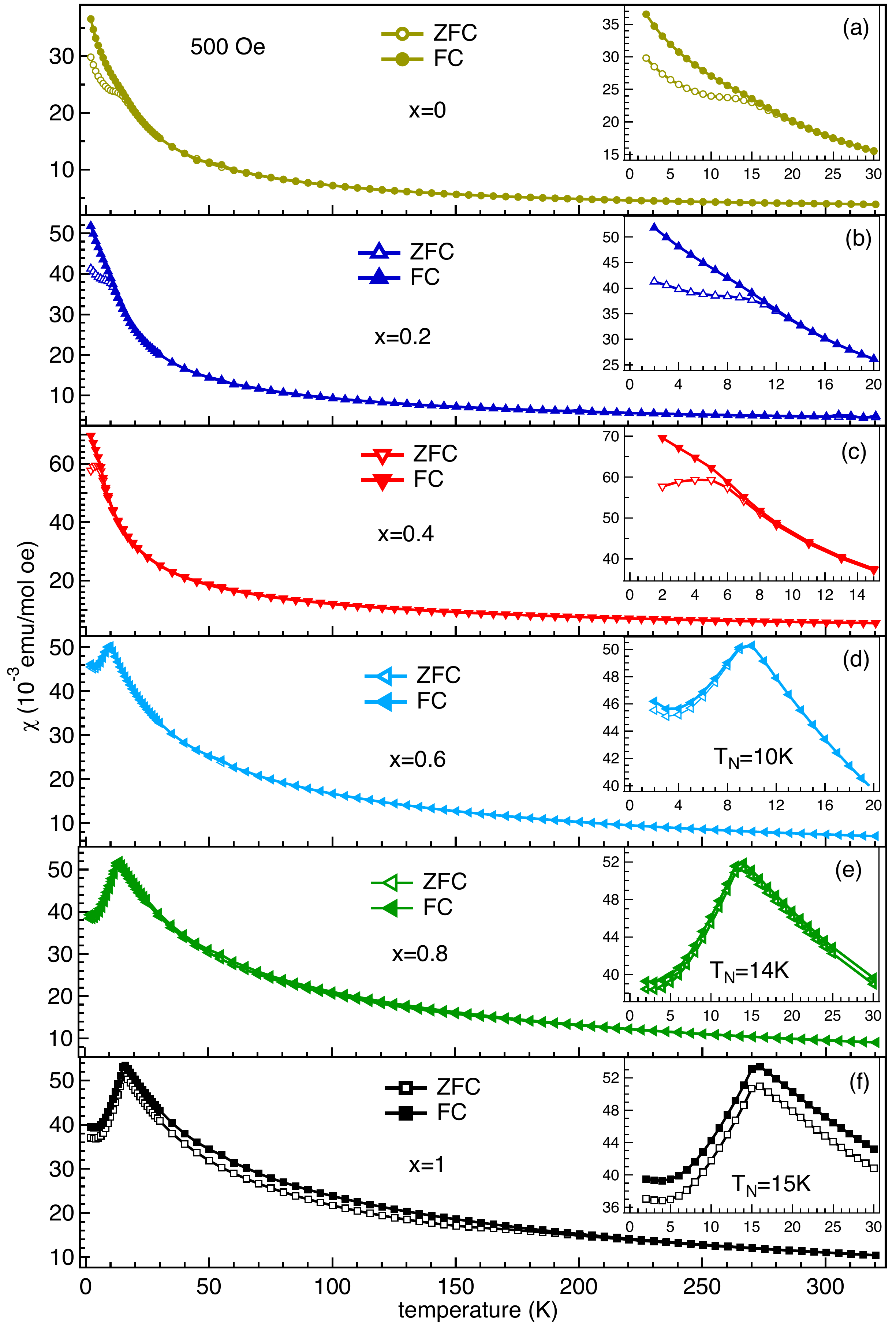}
\caption {(a--f) The magnetic susceptibility of $x=$ 0--1 samples measured in both zero field cooled (ZFC) and field cooled (FC) modes from 2~K to 320~K at 500~Oe, inset in each panel shows the magnification of the low temperature region for the clarity of the presentation.}  
\label{ZFC-FC_all}
\end{figure}

In order to probe the impact of enhancement in the Co$^{2+}$ concentration and hence B-site ordering on the magnetic properties of the Sr$_{2-x}$La$_x$CoNbO$_6$ ($x=$ 0--1) samples, the $\chi-$T measurements were carried out from 2 to 320~K in both zero field cooled (ZFC) and field cooled (FC) modes at 500~Oe applied magnetic field, as shown in Figs.~\ref{ZFC-FC_all}(a--f). Inset in each panel shows the enlarged view of low temperature region to make the transition clearly visible. The clear bifurcation in the ZFC and FC curves of 0$\leqslant x \leqslant$0.4 samples indicates the possibility of spin glass (SG) and/or ferromagnetic interactions in the low temperature regime. For the $x=$ 0 sample, we observe a cusp around 15--20~K, which is consistent with the earlier report, where the author claimed this cusp to be the field independent signature, which along with the absence of aging effect discard the possibility of the low temperature spin-glass behavior in the parent compound \cite{Yoshii_JAC_2000}. The ZFC and FC curves bifurcate below around 20~K, 12~K and 7~K for the $x=$ 0, 0.2 and 0.4 samples, respectively, see Figs.~\ref{ZFC-FC_all}(a--c). Interestingly, a decrease in the bifurcation temperature up to $x=$ 0.4 sample suggests the reduction in the ferromagnetic interactions with the La concentration. This is consistent with the field dependent magnetization (M-H) measurements at lower temperature, discussed later. 

\begin{figure}[h]
\includegraphics[width=3.35in]{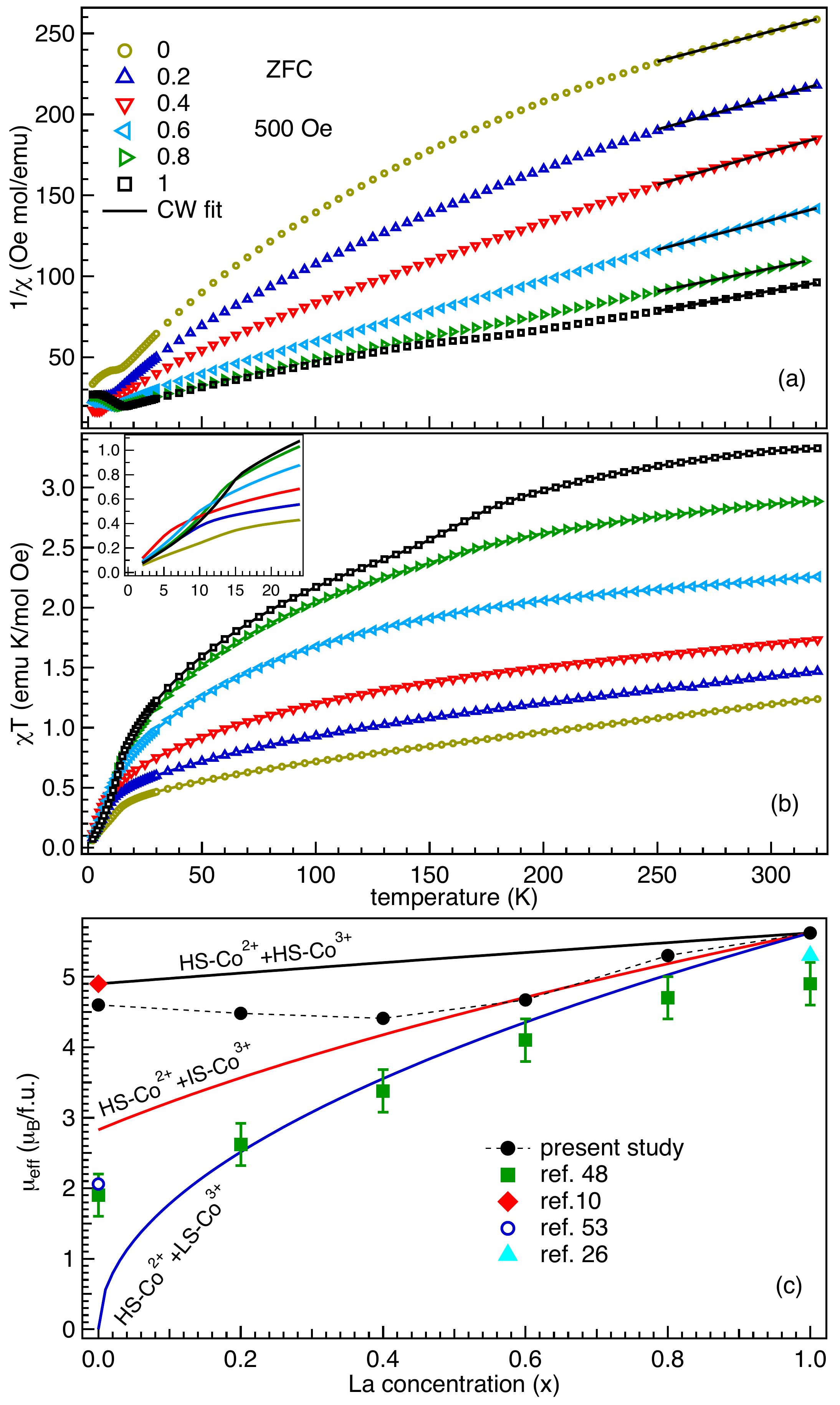}
\caption{Temperature dependent inverse susceptibility (a) and temperature weighted susceptibility ($\chi$T) (b) of Sr$_{2-x}$La$_x$CoNbO$_6$ for $x=$ 0--1 respectively in ZFC mode from 2~K to 320~K in the presence of 500~Oe magnetic field. Inset in (b) show the magnification of the lower temperature regime for the clarity. (c) Comparison of the experimentally calculated effective magnetic moment $\mu_{\rm eff}$ (solid black circles) with that reported in the literature, and expected magnetic moment for the various spin-states of Co$^{3+}$ with increase in La content ($x$), where Co$^{2+}$ is expected and considered to be always in HS state with composition weighted 5.62(1)~$\mu_B$ contribution in the magnetic moment. The error in the $\mu_{\rm eff}$ values of the present study is less than the size of the symbols.}
\label{FC-all}
\end{figure}

Moreover, with further increase in the La concentration i.e., $x \geqslant $ 0.6, we observe an antiferromagnetic ordering at low temperature. More interestingly, the value of transition temperature, T$_{\rm N}$ increases with La concentration i.e., T$_{\rm N}\approx$10, 14 and 16~K for the $x=$ 0.6, 0.8 and 1 samples, respectively, see Figs.~\ref{ZFC-FC_all}(d--f). Notably, the value of T$_{\rm N}$ for the $x=$ 1 sample is in close agreement with the reported in ref.~\cite{Bos_PRB_04}. It is interesting to note that alternating ordering of NbO$_6$ and CoO$_6$ octahedra leads to the antiferromagnetic interaction between two Co ions through 90 and 180$\degree$ Co-O-Nb-O-Co superexchange paths \cite{Bos_PRB_04}. The enhancement in the T$_{\rm N}$ directly support the fact of increase in the B-site ordering in these samples, which is the consequence of increase in the Co$^{2+}$ concentration with La substitution. There is no significant bifurcation observed for the $x=$ 0.6 and 0.8 samples, while for the $x=$ 1 sample the ZFC--FC curves bifurcate below around 180~K [see Fig.~\ref {ZFC-FC_all}(f)]. This suggests the presence of weak ferromagnetic interactions in the $x=$ 1 sample, which is further supported by the M--H measurements at 30~K (discussed later). The presence of small amount of secondary phases in the highly La rich sample, which are out of the detectability limit of the conventional XRD, could be the possible reason for this. Furthermore, Figs.~\ref{FC-all}(a, b) show the temperature dependence of inverse and temperature weighted susceptibility, respectively. A non-linearity in the inverse susceptibility curves in the $x\leqslant$ 0.4 samples is possibly due to the strong crystal field effect in case of Co$^{3+}$. Interestingly, a monotonic enhancement in the magnetic susceptibility is observed with the La substitution due to increment in the temperature independent paramagnetic (TIP) moment owing to the enhancement in the Co$^{2+}$ concentration, having $^4T_1$ ground state term. The Co$^{2+}$ have significantly non-zero TIP term because of the mixing in ground and the excited states due to non-zero second order Zeeman's interaction \cite{Lloret_ICA_08}. Inset of Fig.~\ref{FC-all}(b) show the enlarged view of the low temperature regime, where a reduction in $\chi$T values for $x\geqslant$ 0.6 below T$_{\rm N}$ indicate the domination of AFM interactions over TIP moment.

First we analyze the high temperature magnetization data (250--320~K) using Curie--Weiss (C--W) law, $\chi$ = C/(T-T$_\theta$), where C and T$_\theta$ are Curie-Weiss constant and Curie temperature, respectively. The values of C, T$_\theta$ and calculated $\mu_{\rm eff}$ per formula unit extracted from the linear fitting of inverse susceptibility versus temperature curves [see Fig.~\ref{FC-all}(a)] are given in the table~II. Here, the value of the frustration factor $f$, defined as $f=$ $\mid{T_\theta}\mid$/T$_t$, where T$_t$ is the transition temperature, taken as T$_{\rm N}$ for $x \geqslant $ 0.6 and bifurcation temperature in ZFC and FC curves for $x\leqslant$ 0.4, indicate the highly frustrated spins in all the samples \cite{Bos_PRB_04}. The strong competition between the various possible oxidation and spin-states of Co along with the B-site disorder and complex exchange interaction between them give rise to the frustration in the spins at low temperature, which can strongly influence the magnetic transition temperature, whereas the value of T$_\theta$ is the measure of the strength of the paramagnetic interactions \cite{Bos_PRB_04, Kumaar_PRB_12}. Note that the $\mu_{\rm eff}$ for the $x=$ 1 sample is higher than the spin only magnetic moment ($\mu_s$) even for high spin Co$^{2+}$ (3d$^7$; t$_{2g}^5$e$_g^2$; S=3/2; 3.87~$\mu_B$), indicating the significant orbital contribution in the magnetic moment due the presence of triply degenerate $^4T_1$ ground term. This unquenched orbital magnetic moment, a common phenomenon for six fold coordinated HS Co$^{2+}$ \cite {Viola_CM_03}, indicate a negligible distortion in CoO$_6$ octahedron. In order to estimate the spin-state of Co$^{3+}$ in all the samples, we examine the behavior of paramagnetic moment and calculate the $\mu_{\rm eff}$ per Co$^{3+}$ ion as $\mu_{\rm Co^{3+}}=\sqrt{ (\mu^2_{\rm eff} - x . \mu^2_{\rm Co^{2+}})/(1-x)}$, where $ \mu_{\rm eff}$ is the experimentally calculated effective magnetic moment per formula unit. Here, we consider the composition weighted 5.62(1)~$\mu_B$ contribution from Co$^{2+}$ i.e., $x . \mu_{\rm Co^{2+}}$ in above equation. From the calculated magnetic moment for Co$^{3+}$ (see Table~II), we estimate the spin-states of Co$^{3+}$ in these samples by considering spin only contribution in the magnetic moment of Co$^{3+}$ ions. 

\begin{table}[h]
\centering
\label{table:mu_all}
\caption{Curie-Weiss temperature (T$_\theta$), Curie constant (emu K/mol Oe), paramagnetic effective magnetic moment ($\mu_B$) extracted from the linear fitting of 1/$\chi$ versus T plot at the higher temperature, calculated effective moment of Co$^{3+}$ ($\mu_B$) and frustration factor ($f$) for Sr$_{2-x}$La$_x$CoNbO$_6$ for $x=$ 0--1. The numbers in the parentheses represent the errors in the last digit of the extracted parameters.}

\begin{tabular}{p{1cm}p{1.3cm}p{1.4cm}p{1.4cm}p{1.4cm}p{1.3cm}}
\hline
\hline
$x$	& T$_\theta$ (K) &C& $\mu_{\rm eff}$ & $\mu_{\rm Co^{3+}}$ & $f$\\
\hline

0  & -369(5)& 2.64(2)&4.60(2) &4.60(2) & 18.4(2)\\
0.2 &-230(8) &2.51(4)& 4.48(4) &4.15(5) & 19.2(7)\\

0.4 &-131(3)&2.43(1)& 4.41(1) &3.37(3) & 18.7(4)\\

0.6 &-67(1) &2.72(1)& 4.67(1) &2.67(6) & 6.7(1)\\

0.8 &-69(1) &3.52(1)&5.30(1) & 3.76(8) & 4.9(1)\\

1.0 &-65(2) &3.95(2)&5.62(1)&- &4.3(1)\\
\hline
\hline
\end{tabular}
\end{table}

Note that for the $x=$ 0 sample, Co is present only in 3+ valance state, and the estimated value of  $\mu_{\rm eff}$ [4.60(2)~$\mu_B$/Co$^{3+}$] is slightly lower than the spin only value for HS Co$^{3+}$ (4.90~$\mu_B$/Co$^{3+}$) and higher than the IS (2.83~$\mu_B$/Co$^{3+}$) as well as non-magnetic LS states. Therefore, the presence of mixed LS--HS or IS--HS or LS--IS--HS states are expected with dominant contribution from HS states, as reported in ref.~\cite{Yoshii_JAC_2000}. However, combination of LS and IS can not explain the measured magnetic moment. Further, the magnetic moment ($\mu_{\rm eff}$) decreases with increase in the La content from $x=$ 0 to 0.4, which clearly indicates the spin-state transition of Co$^{3+}$ from HS  to IS/LS states. Now let's consider the $x=$ 0.6 sample where the magnetic moment for Co$^{3+}$ [2.67(6)~$\mu_B$] is slightly less than that of IS state (2.83~$\mu_B$) indicating the possibility of LS--HS or LS--IS or LS--IS--HS models only and discard the presence of the IS--HS model, see Fig.~\ref{FC-all}(c) or Table~II. This spin-state transition of Co$^{3+}$ from HS to LS/IS states can be well understood in terms of the chemical pressure exerted on the atoms due to ionic mismatch, as the smaller size of La$^{3+}$ as compared to Sr$^{2+}$ forces the B-site cations to reduce their size in order to maintain the crystal symmetry ($\bar{d}_{\rm A/A'-O} = \sqrt{2} \bar{d}_{\rm B/B'-O}$). An increase in the concentration of larger sized Co$^{2+}$ than Co$^{3+}$ with $x$ further emphasize this fact. In this case, the Co$^{3+}$ with larger crystal field splitting as compared to Co$^{2+}$ prefer to change from HS state to LS/IS states due to having its smaller ionic size in the latter cases. A similar kind of steric effect is also reported in \cite{Knizek_PRB_12}, where the spin-states of Co$^{3+}$ in LaCo$_{1-x}$Rh$_x$O$_3$ stabilized in HS state by the substitution of larger Rh$^{3+}$ cation at smaller Co$^{3+}$ site. In the present case, an unusual increment in the magnetic moment of Co$^{3+}$ for the $x=$ 0.8 sample needs further investigation using direct probe like x-ray spectroscopy \cite{MerzPRB10}, as the possible errors due to the presence of small amount of oxygen non-stoichiometry and role of orbital contribution in IS and HS Co$^{3+}$ cannot be completely ignored in these samples.

In Fig.~\ref{FC-all}(c), we compare the experimentally calculated effective magnetic moment in the present study using C--W law (solid black circles) with that reported in refs.~\cite{Yoshii_JAC_2000, Bos_PRB_04, Kobayashi_JPSJ_12, Azcondo_Dalton_15} as well as the possible mixture of the different spin-states of Co$^{3+}$ with La content, keeping composition weighted 5.62(1)~$\mu_B$ contribution from Co$^{2+}$ HS states. The upper (black), middle (red) and lower (blue) solid lines show the expected effective magnetic moment for the mixtures of HS Co$^{2+}$ with HS, IS, and LS Co$^{3+}$, respectively. Note that a large difference in the absolute values of the effective magnetic moments is reported \cite{Yoshii_JAC_2000, Bos_PRB_04, Kobayashi_JPSJ_12, Azcondo_Dalton_15}, specially for the $x\leqslant$ 0.4 samples in the present study. This is possibly due to the inclusion of wide non-linear low temperature region in the fitting of inverse susceptibility curves \cite{Kobayashi_JPSJ_12, Azcondo_Dalton_15}, where short-range magnetic correlations are present in these samples, and overestimated values of the TIP contribution \cite{Azcondo_Dalton_15}. For example, the TIP contribution and frustration in the $x\leqslant$ 0.4 samples reported in ref.~\cite{Kobayashi_JPSJ_12} are significantly different as compared with the present study (see Fig.~\ref{FC-all} and Table~II) as well as with ref.~\cite{Yoshii_JAC_2000}. More importantly, we have selected a high temperature paramagnetic region for the fitting as there is strong possibility of the temperature dependent change in the relative population of the different spin states of Co$^{3+}$ due to the strong crystal field effect in the $x\leqslant$ 0.4 samples.
 
 \begin{figure} [h]
\centering
\includegraphics[width=3.4in]{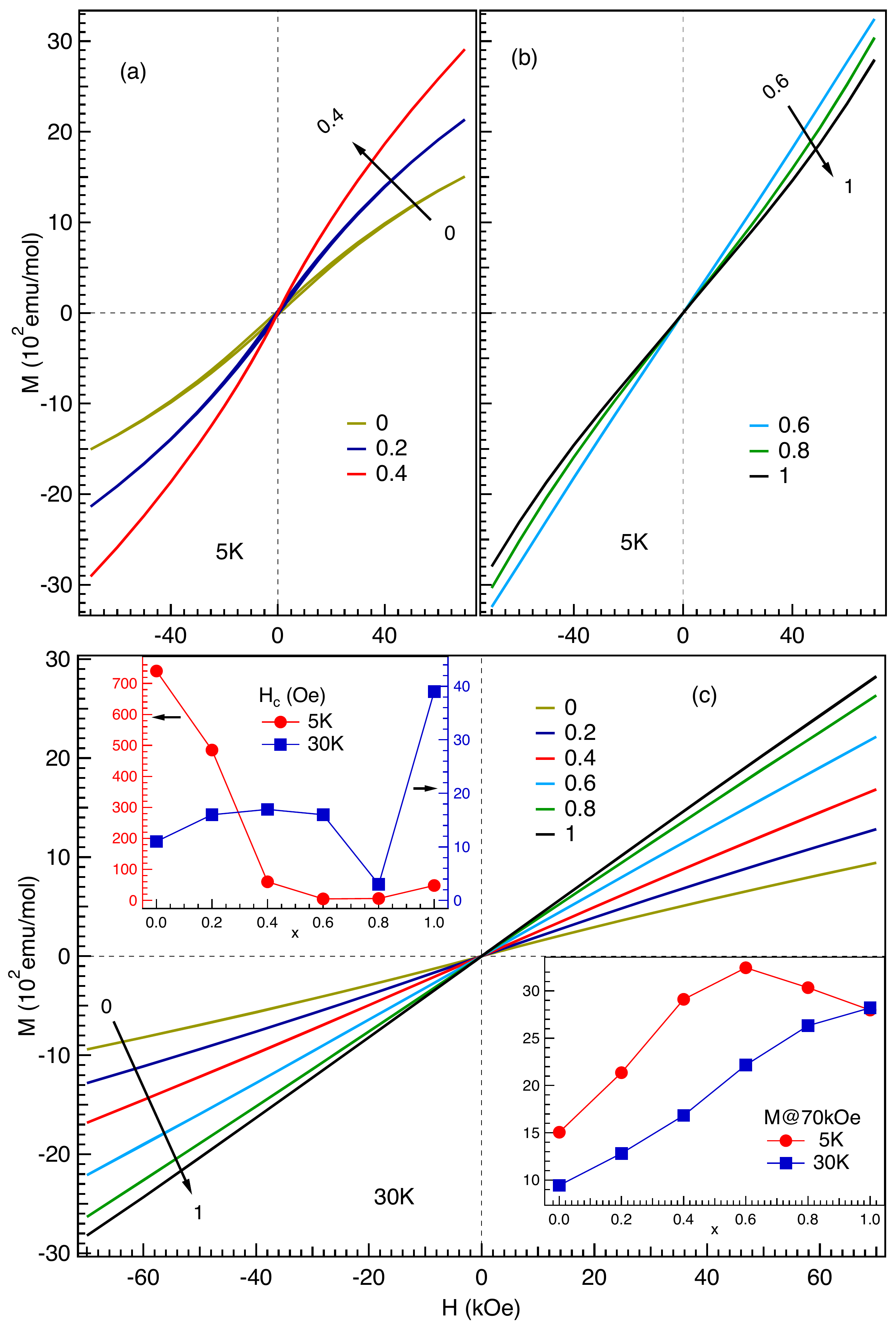}
\caption{Field dependent isothermal magnetization curves of Sr$_{2-x}$La$_x$CoNbO$_6$ for $x=$ 0--0.4 (a) and  0.6--1 measured at 5~K (b) and 0--1 at 30~K (c). Upper and lower insets in (c) show the coercivity and magnetization at 70~kOe, respectively, at 5~K (red solid circles) and 30~K (blue solid squares).} 
\label{MH_all}  
\end{figure}

In order to further understand the complex magnetic behavior in these samples, the isothermal magnetization (M--H) measurements have been performed and shown in the Fig.~\ref{MH_all}, both at 5~K$<$T$_t$ (a, b) and  at 30~K$>$T$_t$ (c), up to $\pm$ 70~kOe applied magnetic field. Upper and lower insets in Fig.~\ref{MH_all}(c) show the coercivity and magnetization at 70~kOe, respectively, at 5~K (red solid circles) and 30~K (blue solid squares) with La concentration. The non-saturating behavior of the M--H curves at 5~K even up to 70~kOe suggests the presence of canted spins in $x\leqslant$0.4 samples at low temperature. On the other hand, considerable values of coercivity and retentivity indicate the presence of small ferromagnetic interactions in these samples. The decrease in coercivity (H$_{\rm C}=$ 740~Oe, 485~Oe and 60~Oe for $x=$ 0, 0.2 and 0.4 samples, respectively) indicates the reduction in the ferromagnetic interactions, which is consistent with the reduction in the bifurcation temperature for $x=$ 0--0.4 samples in $\chi$--T data (see Fig.~\ref{ZFC-FC_all}). The M--H curves for $x=$ 0--0.4 samples at 5~K in Fig.~\ref{MH_all}(a) show a significant increase in the magnetization with $x$. This is due to an increase in TIP contribution from Co$^{2+}$ concentration, which dominates over the effect of decreasing (increasing) ferromagnetic (antiferromagnetic) interactions. Interestingly, for $x\geqslant0.6$ the magnetic moment decreases at 5~K [see Fig.~\ref{MH_all}(b)], which clearly indicates the dominance of the AFM interactions between octahedrally coordinated HS Co$^{2+}$ ions as compared to their TIP contribution. The convex shaped M--H curves at high magnetic field for $x \geqslant$0.6 samples show a weak tendency of the metamagnetism in the system \cite{BollettaPRB18}, which can be further probed by the high field M--H measurements. We have also recorded the magnetic isotherms at 30~K, in order to probe the magnetic interactions above the transition temperature, as shown in Fig.~\ref{MH_all}(c). It is notable that the magnetic moment consistently increase with $x$ due to enhancement of the high moment Co$^{2+}$ concentration and absence of any long range magnetic ordering at this temperature. All the M--H curves in Fig.~\ref{MH_all}(c) show non-saturating behavior with negligible hysteresis (see upper inset) suggesting paramagnetic nature at 30~K, except for the $x=$ 1 sample. 

\begin{figure}[h]
\centering
\includegraphics[width=3.4in]{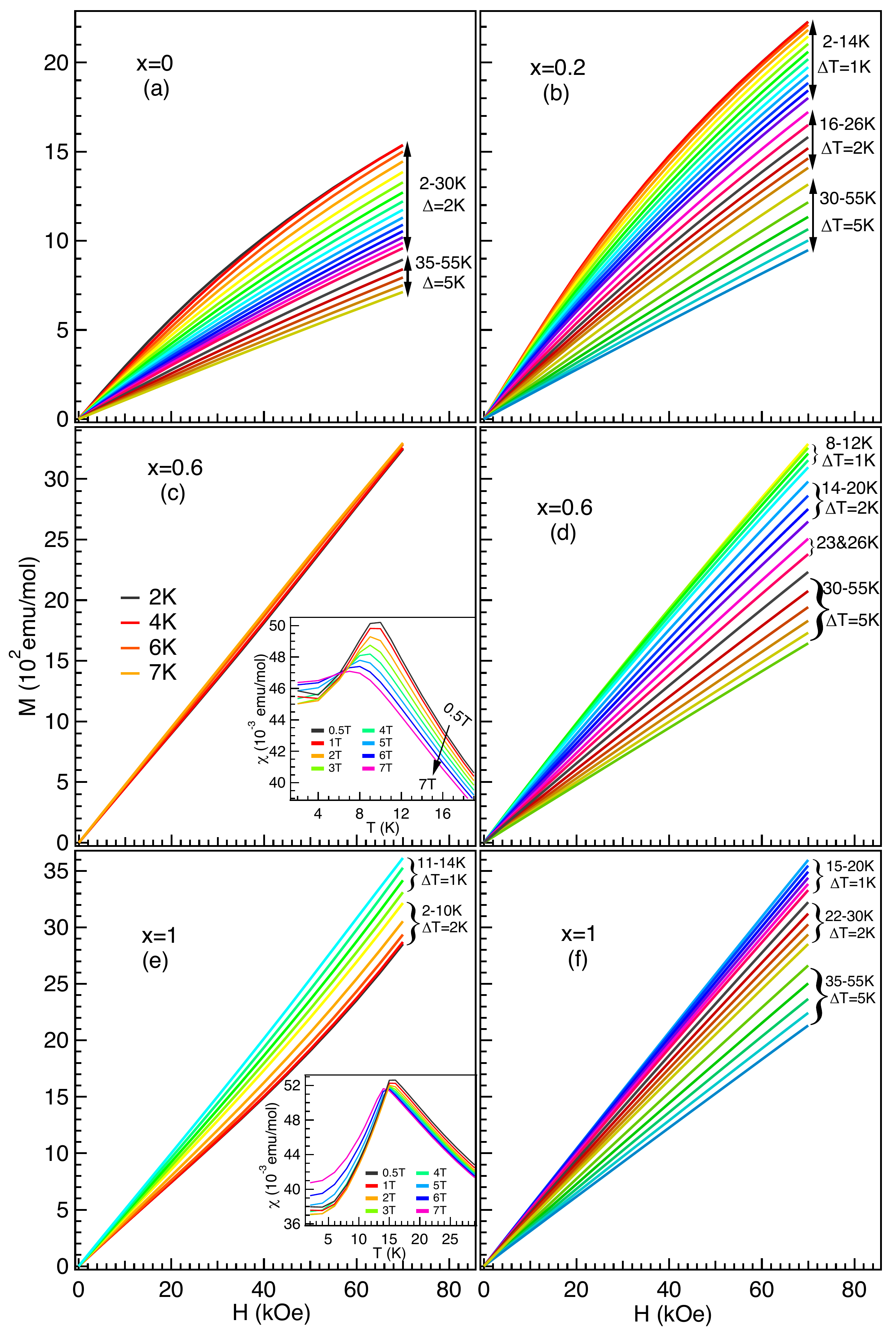}
\caption{Virgin magnetization isotherms for the $x=$ 0 (a), 0.2 (b), 0.6 (c and d) and 1 (e and f) samples from 0 to 70~kOe. Insets in (c) and (e) show susceptibility at different applied magnetic fields extracted from the virgin magnetization isotherms for the $x =$ 0.6 and 1 samples, respectively}.
\label{Fig6_Virgin_All}  
\end{figure}

It is important to emphazise here that our $\chi$--T and M--H measurements clearly demonstrate an evolution of the AFM interactions due to enhancement in the antiferromagnetic HS Co$^{2+}$ concentration with La substitution. At the same time, the presence of small hysteresis for $x\le$ 0.4 samples also indicates the low temperature FM ordering. In order to examine the strong competition between FM and AFM interactions due to the change in spin-state of Co$^{3+}$ and valence state of Co (i.e., from Co$^{3+}$ to Co$^{2+}$) with La substitution, we study the change in the magnetic entropy ($\Delta S$) for the $x=$ 0, 0.2, 0.6 and 1 samples, as it is more sensitive to the complex magnetic interactions, where positive and negative values of $\Delta$S indicate the dominating AFM and FM interactions, respectively. Figs.~\ref{Fig6_Virgin_All}(a--f) show the virgin magnetization isotherms for these samples up to 70~kOe at various temperatures ranging from 2 to 55~K. The magnetic moment values at 2~K and 70~kOe are found to be 1535~emu/mol (0.28~$\mu_B$/f.u.) and 2230~emu/mol (0.40~$\mu_B$/f.u.) for the $x=$ 0 and 0.2 samples, respectively [see Figs.~\ref{Fig6_Virgin_All}(a, b)]. These values are significantly lower than the theoretical saturation moments of free Co$^{3+}$/Co$^{2+}$ ions (M$_s$ = g$_J$J$\mu_B$=6~$\mu_B$), indicating the effect of strong crystal field and/or canting in the spins at low temperature. We observe a slight convex nature and a linear behavior of virgin isotherms below and above T$_{\rm N}$, respectively in the $x=$ 0.6 and 1 samples [see Figs.~\ref{Fig6_Virgin_All}(c--f)]. To understand further we plot the temperature dependent susceptibility at different applied magnetic fields extracted from the virgin magnetization isotherms for the $x=$ 0.6 and 1 samples, respectively, as shown in the insets of Figs.~\ref{Fig6_Virgin_All}(c, e). The susceptibility value at 2~K first decreases with the magnetic field up to $\sim$2~T and then increases up to 7~T, which is the consequence of the weak metamagnetic signature as evident in the M--H measurements [see Fig.\ref{MH_all}(b)], where magnetic moment increase with the relatively slow rate at the lower applied magnetic field and then sharply increases at the higher magnetic fields due to the weak tendency of the spin reorientation \cite{BollettaPRB18}. Further, an increase in the broadening and shift of the transition temperature to the lower value are evident in both the samples [see insets in Figs.~\ref{Fig6_Virgin_All}(c, e)], confirming the AFM interactions. However, both the effects are less prominent in case of the $x=$ 1 sample as compared to the $x=$ 0.6, indicating the enhancement in the strength of AFM interactions with La substitution.

\begin{figure}
\centering
\includegraphics[width=3.4in]{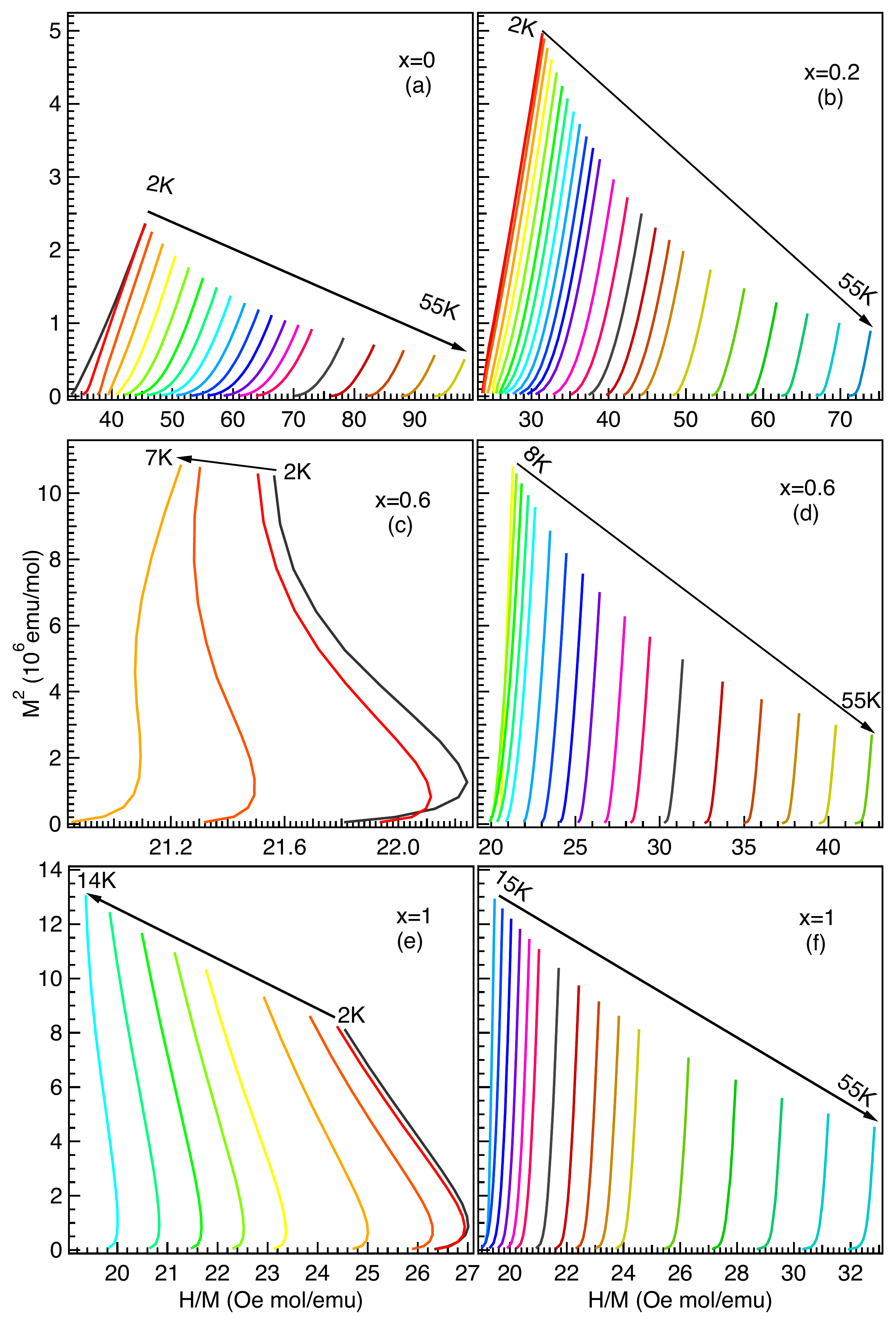}
\caption{Arott plots (M$^2$ versus H/M) for the $x=$ 0 (a), 0.2 (b), 0.6 (c and d) and 1 (e and f) samples.}
\label{Fig7_Arrott_all}  
\end{figure}

\begin{figure*}
\centering
\includegraphics[width=7.0in]{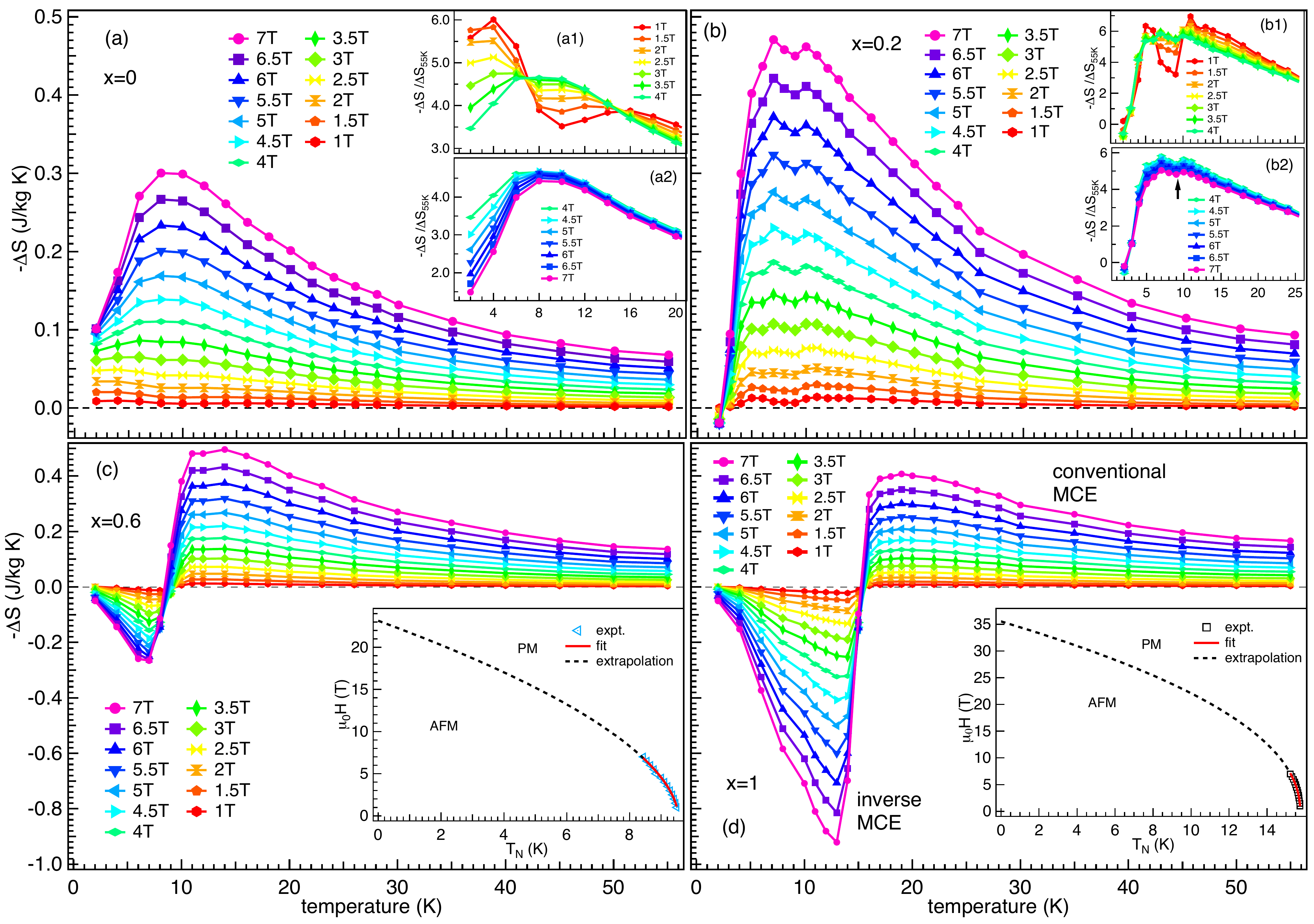}
\caption{Temperature dependent magnetic entropy change at different $\mu_0$$\Delta$H values for $x=$ 0 (a), 0.2 (b), 0.6 (c) and 1 (d) samples. Upper and lower insets in (a) and (b) show the low temperature region of the normalized $\Delta$S with respect to $\Delta$S value at 55K at various $\mu_0$$\Delta$H values for the clear presentation. Insets in (c) and (d) represent the magnetic field evolution of the crossover temperature of $\Delta$S curves for the $x =$ 0.6 and 1 samples, respectively. Red solid line indicate the fit to the data points (see the text) and black dotted line represents the extrapolation of the fit.}
\label{Fig8_entropy_all}  
\end{figure*}

Furthermore, to investigate the nature and strength of the low temperature complex magnetic interactions, in Figs.~\ref{Fig7_Arrott_all}(a--f), we present the Arrott plots (M$^2$ versus H/M), derived from the virgin isotherms, which show a positive intercept on the $x-$ axis indicating the zero spontaneous magnetization \cite{Arrott_PR_57} for the $x=$ 0, 0.2, 0.6 and 1 samples. Also, the presence of the positive curvature for all the samples in the very low field region is the consequence of the random local field as proposed by Aharony and Pytee \cite{Aharony_PRL_80}. For the $x=$ 0 and 0.2 samples, the Arrott curves shift towards higher H/M values with temperature and are almost straight up to ~10K; whereas a change to positive curvature is observed for $>$10~K at low magnetic fields [see Figs.~\ref{Fig7_Arrott_all}(a, b)]. However, a linear behavior with positive slope i.e. curves are not vertically straight up to 55~K in the high field region suggests the presence of short-range magnetic correlations well above the bifurcation temperature. We note here that, for the $x=$ 0.6 and 1 samples, the Arrott curves move towards lower H/M values up to a certain temperature (T$_{\rm N}$) and then shift to higher side with further increase in temperature up to 55~K, see Figs.~\ref{Fig7_Arrott_all}(c--f). The M$^2$ curves show buckling like ``S" shape, which is the signature of metamagnetic transition in these samples. We observed the negative (positive) slopes of the Arrott plots in moderate (high) magnetic fields, which suggests that the field induced phase transition in these samples is complex in nature \cite{Banerjee_PL_64, Midya_PRB_11}. Therefore, it is vital to perform further analysis of magnetic entropy change, which is considered to be more accurate to understand the complex magnetic interactions/behavior and magnetocaloric effect.

Now we move to the discussion of the change in the magnetic entropy ($\Delta$S) induced by change in the magnetic field ($\mu_0$$\Delta$H) as a function of temperature, which is estimated using the following Maxwell's thermodynamic relation \cite{Phan_JMMM_07}.
\begin{eqnarray} 
 \Delta S (T,H)=\mu_0\int_0^H\bigg (\frac{\partial M(T,H)} {\partial T}\bigg)_H dH
\end{eqnarray}
In Figs.~\ref{Fig8_entropy_all}(a--d) we plot the temperature dependent $-\Delta$S at various magnetic field change ($\mu_0$$\Delta$H) values from 1 to 7~T for the $x=$ 0, 0.2, 0.6 and 1 samples, respectively. In addition, the insets in Figs.~\ref{Fig8_entropy_all}(a, b) show the low temperature region after normalizing the $\Delta$S with its value at 55~K i.e., $-\Delta$S(T, H)/$\Delta$S$_{\rm 55K}$(H) for the $x =$ 0 and 0.2 samples, respectively. The overall negative values of $\Delta$S for the $x=$ 0 and 0.2 samples indicate the resultant ferromagnetic interactions i.e., conventional magnetocaloric effect (MCE) [see Figs.~\ref{Fig8_entropy_all}(a, b)]. More interestingly, the entropy curves [$-\Delta$S(T, H)/$\Delta$S$_{\rm 55K}$(H) versus temperature] for the $x=$ 0 and 0.2 samples show a local minima peak at 9--10~K at lower $\mu_0$$\Delta$H values, see clearly in the insets (a1) and (b1) of Figs.~\ref{Fig8_entropy_all}(a, b), which indicates the competition between FM--AFM interactions \cite{Kumar_PRB_08}. However, the $\Delta$S values should change from negative to positive when the resultant dominating AFM interactions are present in the system, i.e., inverse MCE is expected due to an increase in the magnetic moment with the temperature. Here, this minima position shifts towards the lower temperature and eventually disappear with increasing $\mu_0$$\Delta$H for the $x=$ 0 sample [see inset (a2) of Fig.~\ref{Fig8_entropy_all}(a)]; however, the minima is still visible in the $x=$ 0.2 sample even at 7~T [see arrow in the inset (b2) of Fig.~\ref{Fig8_entropy_all}(b)] suggesting the enhancement in the AFM interactions with La substitution. On the other hand, the weak antiferromagnetism can be suppressed at high magnetic field where all the spins align in the easy direction, i.e. mixed FM--AFM interactions showing the metamagnetic transitions have been reported in literature \cite{Kumar_PRB_08, Midya_PRB_11, Emre_PRB_08}. Furthermore, the peak value of the entropy change (i.e., --$\Delta S_{max}$) at 7~T found to be $\approx$0.30 and 0.45~J/Kg K for $x =$ 0 and 0.2 samples, respectively, indicating a significant increase with La substitution ($x$). 

More interestingly, with further increase in the La concentration, we observe a negative to positive crossover in the $\Delta$S curves for the $x=$ 0.6 and 1 samples [see Figs.~\ref{Fig8_entropy_all}(c, d)], at T$_{\rm N}\approx$ 9~K and 15~K, respectively. This indicate the well known transition from conventional (at T$>$T$_{\rm N}$) to inverse (at T$<$T$_{\rm N}$) magnetocaloric effect, as observed for the AFM materials \cite{Krenke_NM_05, Biswas_JAP_13}. We observe a significant increase in the positive and a slight decrease in negative values of the $\Delta$S$_{max}$ for the $x=$ 1 sample as compared to the $x=$ 0.6 sample [Fig.~\ref{Fig8_entropy_all}(d)]. However, a shift in the crossover temperature to the lower value has been observed for both the $x=$ 0.6 and 1 samples with increase in the magnetic field as a consequence of the decay of the AFM coupling. Insets of Fig.~\ref{Fig8_entropy_all}(c) and (d) show the $\mu_0$H versus T$_{\rm N}$ phase diagram for the $x=$ 0.6 and 1 samples, respectively, where the crossover temperature of the entropy is considered to be the transition temperature (T$_{\rm N}$). The temperature and field dependence of the decay of the AFM interactions can be estimated by fitting with the equation H = H$_0$(1-T/T$_{\rm N}$)$^\psi$ as represented by a solid red line, where H$_0$ is the critical magnetic field required to break the AFM interactions at 0~K. The H$_0$ and T$_{\rm N}$ are extracted from the extrapolation of the curves to 0~K temperature and 0~T magnetic field as shown by the black dotted line. We obtained H$_0=$ 23(1)~T and 35(2)~T, T$_{\rm N}=$ 9.5~K and 15.7~K, and $\psi$ = 0.56(3) and 0.47(3) for the $x=$ 0.6 and 1 samples, respectively. The high value of the critical magnetic field, H$_0$ and transition temperature, T$_{\rm N}$ for the $x=$ 1 sample as compared to the $x=$ 0.6 further support the fact of the enhancement in the AFM coupling with the La concentration, as observed in the $\chi$--T measurements. Moreover, a non-zero value of the magnetic entropy up to well above the T$_t$ is possibly due to the presence of magnetic interactions resulting from the frustrated spins, as also evident from the large value of frustration parameter, see Table~II. In addition, the crossover in the temperature dependent $\Delta$S curves with the significant  positive and negative values suggest that these materials can be potential candidates for magnetic refrigerators as well as heat pumps \cite {Egolf_Conf_05}.

Further, we estimate the power dependence of $\Delta$S on the magnetic field i.e., $\Delta$S $\propto$ H$^n$, where $n$ is the local exponent of the entropy change. In case of ferromagnetic materials, it has been reported that $n$ attains its value 1 and 2 below and above the transition temperature, respectively \cite{Franco_JAP_06}. Whereas, at the transition temperature, the $n$ can be expressed in terms of the critical exponents as $n$(T$_{\rm trans}$)=1 + (1-$\beta^{-1}$)$\delta^{-1}$, where $\beta$ is the temperature dependence of the magnetization in the zero magnetic field (spontaneous magnetization) i.e. M $\propto$ (T$_{\rm C}-$T)$^\beta$ for T$<$T$_{\rm C}$ and $\delta$ is the field dependence of the magnetization at the transition temperature i.e M $\propto$ H$^{1/\delta} $\cite{Arrott_PRL_67, Franco_JAP_06}. The local value of $n$ at each temperature and magnetic field can be calculated using the following equation
\begin{eqnarray} 
n (H,T)=\frac{d (ln \mid \Delta S \mid)} {d (ln H)}
\label{n(H,T)}  
\end{eqnarray}
\begin{figure}
\centering
\includegraphics[width=3.5in]{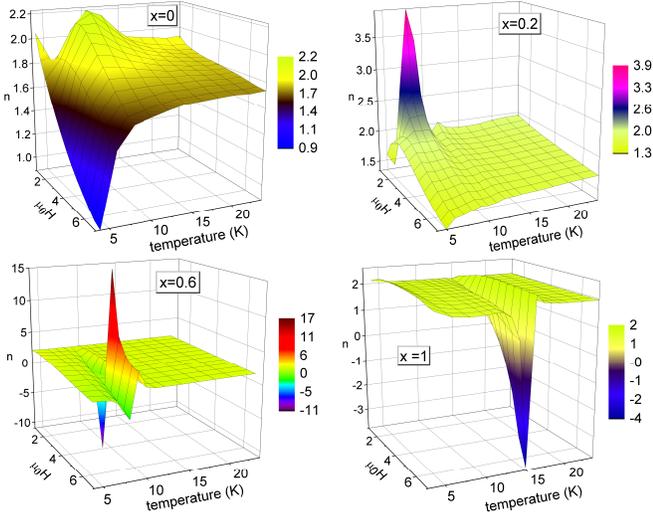} 
\caption{Temperature and magnetic field dependence of exponent $n$ for the $x =$ 0, 0.2, 0.6 and 1 samples.} 
\label{Fig_10_n_3D}  
\end{figure}
Figs.~\ref{Fig_10_n_3D}(a--d) show the 3D plots of the temperature and field dependent $n$ for the $x=$ 0, 0.2, 0.6 and 1 samples. It can be clearly observed that $n$ is less than 2 for the $x=$ 0 and 0.2 samples below the bifurcation temperature in ZFC--FC curves, except an overshoot of $n>$ 2 around 9--10~K in the low magnetic field regions ($<$2~T), where the entropy curves show the strong field dependent behavior [see insets (a1, b1) of Fig.~~\ref{Fig8_entropy_all}]. Whereas, in case of the $x=$ 0.6 and 1 samples, $n$ is approximately 2 below T$_{\rm N}$ and then shows a characteristic peak at the IMCE--CMCE crossover followed by $n=$ 2 in the higher temperature regime (PM) as a consequence of the C--W law. Recently, Law {\it et al.} have proposed a quantitative general criterion based on the magnetocaloric effect to understand the order of magnetic phase transition and successfully applied to various compounds including cobaltites having AFM--FM and FM--PM type transitions \cite{Law_nature_18}. The authors showed that the overshoot in $n$ above 2 is a signature of the first order AFM--FM transition in GdBaCo$_2$O$_{6-\delta}$ \cite{Law_nature_18}. However, in the present study the values of exponent $n$ are strongly field dependent, for example a sudden transition around 4~T can be seen for the $x=$ 0.6 sample at around T$_{\rm N}$. Here our combined analysis of Arrott curves, and temperature and field dependence of exponent $n$ suggest a first order transition at low magnetic fields and second order at higher fields \cite{SarkarPRB08, Midya_PRB_11}. In addition, the temperature dependent specific heat measurements at different applied magnetic fields would be useful to further understand the complex magnetic phase transitions in these samples.

 \begin{figure}
\centering
\includegraphics[width=3.45in]{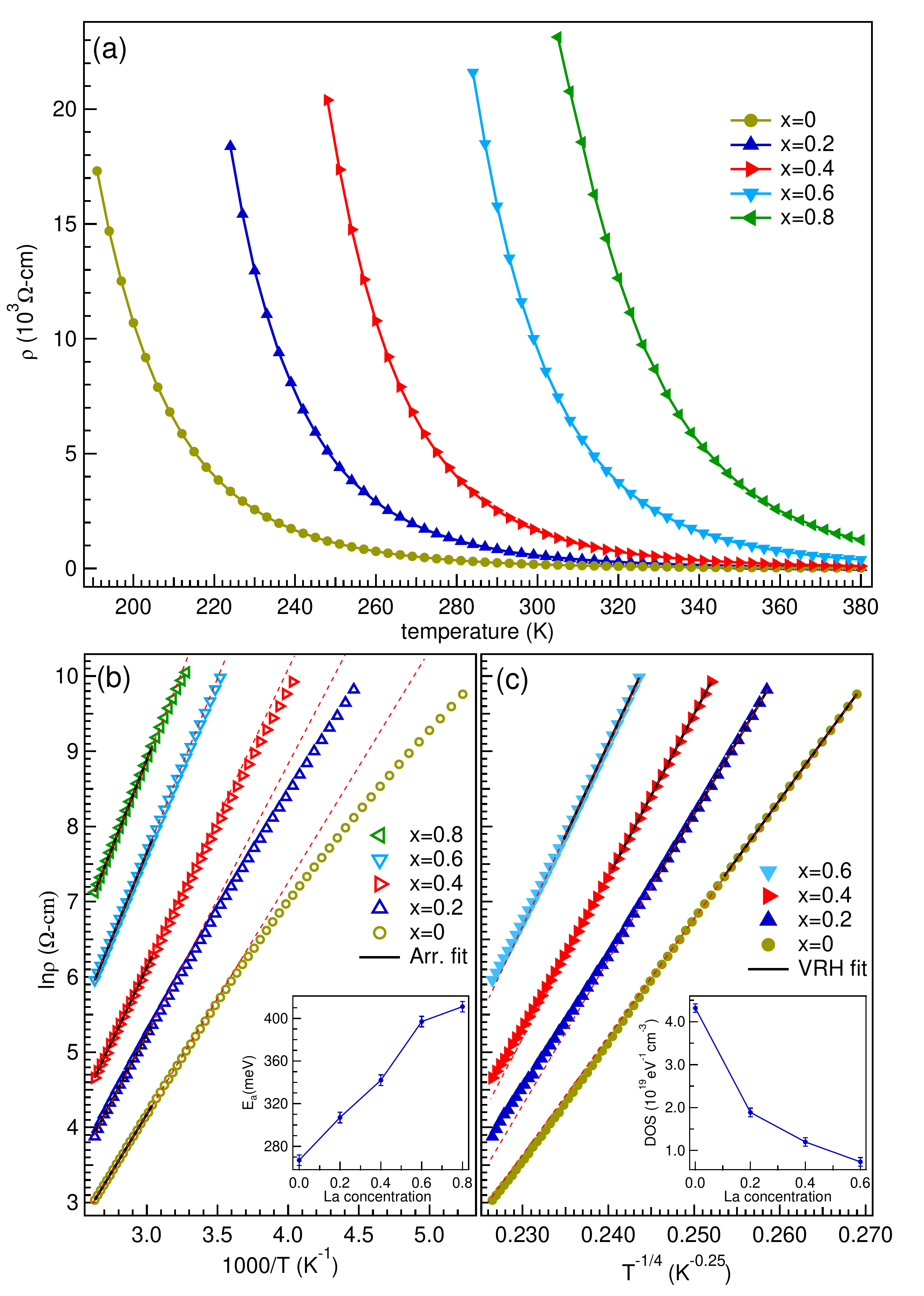}
\caption{(a) Temperature dependent electrical resistivity of  Sr$_{2-x}$La$_x$CoNbO$_6$  for $x$ = 0--0.8. (b) Arrhenius plot of ln($\rho$) versus 1000/T and (c) VRH plot of ln($\rho$) versus T$^{-1/4}$. Continuous black lines show the linear fitting of high (low) temperature data and dashed red line show the extrapolation of the linear fitting in the lower (higher) temperature regime. Inset of (b) and (c) represent the variation in the activation energy and DOS respectively, with La concentration.} 
\label{RT_all}  
\end{figure}

Finally we investigate the transport mechanism in Sr$_{2-x}$La$_x$CoNbO$_6$ samples as La substitution converts Co$^{3+}$ to Co$^{2+}$ and expected to increase resistivity significantly. In order to probe this effect, temperature dependent resistivity ($\rho$--T) measurements were performed for the $x=$ 0--0.8 samples and shown in Fig.~\ref{RT_all}(a). The negative temperature coefficient of resistivity indicates the semiconducting/insulating nature of all the samples in the measured temperature range, which is consistent with ref.~\cite{XU_JCSJ_16} for the $x=$ 0 sample. We observe an evolution of strong insulating phenomenon with La substitution i.e. at higher Co$^{2+}$ concentration. This behavior at higher temperature can be understood by the Arrhenius model, which gives an estimation of the energy required for charge carriers to take part in the conduction mechanism. The conductivity (resistivity) can be described by the Arrhenius equation as given below:
\begin{eqnarray}
 \rho(T)= \rho(0)exp(E_a/k_BT) 
\end{eqnarray}
where $\rho$(0) and E$_a$ are pre-exponential factor and activation energy, respectively. In order to extract the activation energy, ln($\rho$) versus 1000/T (from 330 to 380~K) curves are shown in Fig.~\ref{RT_all}(b) where their slope is used to calculate the activation energy (E$_a$) for the $x=$ 0--0.8 samples. The E$_a$ values are plotted in the inset of Fig.~\ref{RT_all}(b), which show increasing trend with La substitution and consistent with resistivity behavior in Fig.~\ref{RT_all}(a). However, the conduction mechanism significantly deviates in lower temperature regime and can be explained by the Mott variable range hopping (VRH) model due to the localized charge carriers. The VRH model for three dimensional conduction describes the temperature dependence of resistivity as \cite{Paul_PRL_73}
\begin{eqnarray}
 \rho(T)= \rho(0)exp(T_0/T)^{1/4}
\end{eqnarray}
where T$_0$ is characteristic temperature given as, T$_0$ = 18/k$_B$N(E$_F$)L$^3$ where N(E$_{\rm F}$) and L are the localized density of states (DOS) and localization length, respectively. The characteristic temperature, T$_0$ is calculated by the linear fitting of ln($\rho$) versus T$^{-1/4}$ plot in the lower temperature region as shown in Fig.~\ref{RT_all}(c). The DOS around the Fermi level can be calculated from the T$_0$ value by taking average Co--O bond length as localization length ($\approx$2~$\rm \AA$). The inset in Fig.~\ref{RT_all}(c) clearly confirms that the density of localized states decreases indicating a reduction in the conductivity with La substitution.
 
Here we note that the drastic decrease in the conductivity (increase in resistivity) at higher La concentration is demonstrated by increase in the activation energy and reduction in density of state near the Fermi level [see insets in Figs.~\ref{RT_all}(b, c)], which can be understood in terms of reduction in the oxidation state of Co ions. In the transition metal oxides, electronic conduction is mainly governed by the exchange mechanism mediated by the transition metal atoms i.e., B-site cations in the double perovskite oxides \cite{XU_JCSJ_16, Tong_PRB_04}. For low La concentration samples, Co and Nb atoms are randomly occupied and hence electronic conduction can mainly take place through Co$^{3+}$--O$^{2-}$--Co$^{3+}$ path. However, at higher La concentration, Co$^{3+}$ converted to Co$^{2+}$, which is the lowest oxidation state of Co and hence suppress this conduction path resulting in the drastic reduction in the electronic conductivity \cite{Viola_CM_03}. Further the enhancement in the B-site ordering in case of La rich samples leads to the evolution of longer Co$^{2+/3+}$--O$^{2-}$--Nb$^{5+}$--O$^{2-}-$Co$^{2+/3+}$ conduction channel, which can also be responsible for this notable reduction in the electronic conductivity with La substitution. 

\section{\noindent ~Conclusions}

We have systematically tuned the structural, magnetic and transport properties through the substitution of Sr$^{2+}$ by non-magnetic La$^{3+}$ ions in Sr$_{2-x}$La$_x$CoNbO$_6$. The Rietveld refinements of XRD data show the phase transformation from tetragonal ($I4/m$) to monoclinic ($P2_1/n$) for the $x \geqslant$ 0.6 samples. An evolution of the superlattice reflections in the XRD patterns and new Raman active modes confirm the enhancement in the B-site ordering with increase in La concentration. The field dependent magnetization measurements shows weak ferromagnetic interactions for the $x\le$ 0.4 samples, while further increase in La concentration give a slight convex like M--H curves, indicating the presence of weak metamagnetism in the samples at 5~K. The magnetic susceptibility data show bifurcation between ZFC and FC curves for the $x\le$ 0.4 sample, whereas a clear AFM ordering appears with T$_{\rm N}=$ 9-15~K for the $x>$ 0.4 samples. Interestingly, the Curie-Weiss behavior of paramagnetic susceptibility reveals the stabilization of Co$^{2+}$ in HS state and a spin-state transition in Co$^{3+}$ from HS to IS/LS states with La substitution. Moreover, our analysis of magnetic entropy change suggests a crossover from conventional to inverse magnetocaloric effect and explicitly confirms an increase in the strength of the AFM interactions. We observe that the temperature dependent resistivity increases significantly, which can be explained by Arrhenius and VRH models. The resistivity behavior is consistent as La substitution converts Co$^{3+}$ to Co$^{2+}$, which suppresses the conduction path. 

\section{\noindent ~Acknowledgments}

This work is supported by SERB-DST through Early Career Research (ECR) Award (project reference no. ECR/2015/000159) to RSD. AK acknowledges the UGC for the fellowship. We thank physics department at IIT Delhi for providing the research facilities: XRD, Raman Spectrometer (UFO scheme), PPMS EVERCOOL-II and SQUID. RSD also acknowledges BRNS for a high temperature furnace (used for sample preparation), procured from the grant of DAE Young Scientist Research Award (project sanction No. 34/20/12/2015/BRNS).


\begin{thebibliography}{99}

\bibitem{Raccah_PR_67} P. M. Raccah, and J. B. Goodenough, First-order localized-electron $\rightleftarrows$ collective-electron transition in LaCoO$_3$, Phys. Rev. {\bf 155}, 932 (1967).

\bibitem{Bhide_PRB_72} V. G. Bhide, D. S. Rajori, G. R. Rao, and C. N. R. Rao, M$\ddot{\rm o}$ssbauer studies of the high-spin-low-spin equilibria and the localized-collective electron transition in LaCoO$_3$, Phys. Rev. B {\bf 6}, 1021 (1972).

\bibitem{Asai_PRB_89} K. Asai, P. Gehring, H. Chou, and G. Shirane, Temperature-induced magnetism in LaCoO$_3$, Phys. Rev. B {\bf 40}, 10982 (1989).

\bibitem{Zhuang_PRB_98} M. Zhuang, W. Zhang, and N. Ming, Competition of various spin states of LaCoO$_3$, Phys. Rev. B {\bf 57}, 10705 (1998).

\bibitem{Yan_PRB_04} Y.-Q. Yan, J.-S. Zhou, and J. B. Goodenough, Bond-length fluctuations and the spin-state transition in LCoO$_3$ (L = La, Pr, and Nd),
Phys. Rev. B {\bf 69}, 134409 (2004).

\bibitem{Korotin_PRB_96} M. A. Korotin, S. Y. Ezhov, I. V. Solovyev, V. I. Anisimov, D. I. Khomskii, and G. A. Sawatzky, Intermediate-spin state and properties of LaCoO$_3$.
Phys. Rev. B {\bf 54}, 5309 (1996).

\bibitem{Chainani_PRB_92} A. Chainani, M. Mathew, and D. D. Sharma, Electron-spectroscopy study of the semiconductor-metal transition in La$_{1-x}$Sr$_x$CoO$_3$,
Phys. Rev. B {\bf 46}, 9976 (1992).

\bibitem{Saitoh_PRB_97} T. saitoh, T. Mizokawa, A. Fujimori, M. Abbate, Y. Takeda, and M. Takano, Electronic structure and temperature-induced paramagnetism in LaCOO$_3$. 
Phys. Rev. B {\bf 55}, 4257 (1997).

\bibitem{Troyanchuk_PSS_05} I. O. Troyanchuk, D. V. Karpinsky, and R. Szymczak, Possible ferromagnetic interactions between IS Co$^{3+}$ ions in Nb doped cobaltites,
Phys. Status Solidi B {\bf 242}, R49 (2005).

\bibitem{Yoshii_JAC_2000} K. Yoshii, Structural and magnetic properties of the perovskites Sr$_{n+1}$(Co$_{0.5}$Nb$_{0.5}$)$_n$ O$_{3n+1}$ (n = $\infty$, 2 and 1), J. Alloy Compd. {\bf 307}, 119 (2000). 

 \bibitem{Galasso_JPC_62} F. Galasso and W. Darby, Ordering of the octahedrally co\"ordinated cation position in the perovskite structure, J. Phys. Chem. {\bf 66}, 131 (1962).

 \bibitem{Anderson_SSC_93} M. T. Anderson, K. B. Greenwood, G. A. Taylor, and K. R. Poeppelmeier, B-cation arrangements in double perovskites, Prog. Solid St. Chem. {\bf 22}, 197 (1993).

 \bibitem{King_JMC_10} G. King and P. M. Woodward, Cation ordering in perovskites, J. Mater. Chem. {\bf 20}, 5785 (2010).

\bibitem{Vasala_SSC_15} S. Vasala and M. Karppinen, A$_2$B$^\prime$B$^{\prime\prime}$O$_6$ perovskites: A review, Prog. Solid St. Chem. {\bf 43}, 1 (2015).

\bibitem{Jung_PRB_07} A. Jung, I. Bonn, V. Ksenofontov, M. Panth\"ofer, S. Reiman, C. Felser, and W. Tremel, Effect of cation disorder on the magnetic properties of Sr$_2$Fe$_{1-x}$Ga$_x$ReO$_6$ (0\textless$x$\textless0.7) double perovskites, Phys. Rev. B {\bf 75}, 184409 (2007). 

\bibitem{Bos_PRB1_04} J. -W. G. Bos and J. P. Attfield, Control of antisite disorder, magnetism, and asymmetric doping effects in (La$_{1+x}$Ca$_{1-x}$)CoRuO$_6$ double perovskites, Phys. Rev. B {\bf 69}, 094434 (2004).

\bibitem{Narayanan_PRB_10} N. Narayanan, D. Mikhailova, A. Senyshyn, D. M. Trots, R. Laskowski, P. Blaha, K. Schwarz, H. Fuess, and H. Ehrenberg, Temperature and composition dependence of crystal structures and magnetic and electronic properties of double perovskites La$_{2-x}$Sr$_x$CoIrO$_6$ ($0\leqslant x\leqslant$2), Phys. Rev. B {\bf82}, 024403 (2010).

\bibitem{Sarma_PRL_07} D. D. Sarma, S. Ray, K. Tanaka, M. Kobayashi, A. Fujimori, P. Sanyal, H. R. Krishnamurthy, and C. Dasgupta, Intergranular magnetoresistance in Sr$_2$FeMoO$_6$ from a magnetic tunnel barrier mechanism across grain boundaries, Phys. Rev. Lett. {\bf 98}, 157205 (2007).

\bibitem{Meneghini_PRL_09} C. Meneghini, S. Ray, F. Liscio, F. Bardelli, S. Mobilio, and D. D. Sarma, Nature of ``disorder" in the ordered double perovskite Sr$_2$FeMoO$_6$, Phys. Rev. Lett. {\bf 103}, 046403 (2009).

\bibitem{Yin_EES_19} W. -J. Yin, B. Weng, J. Ge, Q. Sun, Z. Li, and Y. Yan, Oxide perovskites, double perovskites and derivatives for electrocatalysis, photocatalysis, and photovoltaics, Energy Environ. Sci. {\bf 12}, 442 (2019).

\bibitem{Kangsabanik_PRM_18} J. Kangsabanik, V. Sugathan, A. Yadav, A. Yella, and A. Alam, Double perovskites overtaking the single perovskites: A set of new solar harvesting materials with much higher stability and efficiency, Phys. Rev. Mater. {\bf 2}, 055401(2018).

\bibitem{Huang_CM_09} Y. -H. Huang, G. Liang, M. Croft, M. Lehtim\"aki, M. Karppinen, and J. B. Goodenough, Double-perovskite anode materials Sr$_2$MMoO$_6$ (M = Co, Ni) for solid oxide fuel cells, Chem. Mater. {\bf 21}, 2319 (2009).

\bibitem{Yoo_RSC_14} S. Yoo, J. Kim, S. Y. Song, D. W. Lee, J. Shin, K. M. Ok, and G. Kim, Structural, electrical and electrochemical characteristics of La$_{0.1}$Sr$_{0.9}$Co$_{1-x}$Nb$_x$O$_{3-\delta}$ as a cathode material for intermediate temperature solid oxide fuel cells, RSC Adv. {\bf 4}, 18710 (2014). 

\bibitem{Kobayashi_nature_98} K. -I. Kobayashi, T. Kimura, H. Sawada, K. Terakura, and Y. Tokura, Room-temperature magnetoresistance in an oxide material with an ordered double-perovskite structure, Nature (London) {\bf 395}, 677 (1998).

\bibitem{Sun_AM_18} Q. Sun, J. Wang, W. -J. Yin, and Y. Yan, Bandgap engineering of stable lead-free oxide double perovskites for photovoltaics, Adv. Mater {\bf 30}, 1705901 (2018).

\bibitem{Bos_PRB_04} J. -W. G. Bos and J. P. Attfield, Magnetic frustration in (LaA)CoNbO$_6$ (A=Ca, Sr, and Ba) double perovskites, Phys. Rev. B {\bf 70}, 174434 (2004).

\bibitem{Ding_PRB_19} X. Ding, B. Gao, E. Krenkel, C. Dawson, J. C. Eckert, S. -W. Cheong, and V. Zapf, Magnetic properties of double perovskite $Ln_2$CoIrO$_6$ ($Ln=$ Eu, Tb, Ho): Hetero-tri-spin 3d-5d-4f systems, Phys. Rev. B {\bf 99}, 014438 (2019).

\bibitem{Haripriya_PRB_19} G. R. Haripriya, C. M. N. Kumar, R. Pradheesh, L. M. Martinez, C. L. Saiz, S. R. Singamaneni, T. Chatterji, V. Sankaranarayanan, K. Sethupathi, B. Kiefer, and H. S. Nair, Contrasting the magnetism in La$_{2-x}$Sr$_x$FeCoO$_6$ ($x=$ 0, 1, 2) double perovskites: The role of electronic and cationic disorder, Phys. Rev. B {\bf99}, 184411 (2019).

\bibitem{Mabbs_Dover_08} F. E. Mabbs and D. J. Machin (Eds.), Magnetism and transition metal complexes, Chapman and Hall Ltd., London (1973).

\bibitem{Lloret_ICA_08} F. Lloret, M. Julve, J. Cano, R. Ruiz-Garc\'ia, and E. Pardo, Magnetic properties of six-coordinated high-spin Co(II) complexes: Theoretical background and its application, Inorg. Chem. Acta {\bf 361}, 3432 (2008). 

 \bibitem{Viola_CM_03} M. C. Viola, M. J. Mart\'inez-Lope, J. A. Alonso, J. L. Mart\'inez, J. M. D. Paoli, S. Pagola, J. C. Pedregosa, M. T. Fern\'andez-D\'iaz, and R. E. Carbonio, Structure and magnetic properties of Sr$_2$CoWO$_6$: An ordered double perovskite containing Co$^{2+}$(HS) with unquenched orbital magnetic moment, Chem. Mater. {\bf 15}, 1655 (2003). 

\bibitem{Bos_CM_04} J. -W. G. Bos and J. P. Attfield, Structural, magnetic, and transport properties of  (La$_{1+x}$Sr$_{1-x}$)CoRuO$_6$ double perovskites, Chem. Mater. {\bf 16}, 1822 (2004).

\bibitem{Yuste_DT_11} M. Yuste, J. C. P\'erez-Flores, J. R. de Paz, M. T. Azcondo, F. Garc\'ia-Alvarado, and U. Amador, New perovskite materials of the La$_{2-x}$Sr$_x$CoTiO$_6$ series, Dalton Trans.  {\bf 40}, 7908 (2011).

\bibitem{Lee_PRB_18} M. -C. Lee, S. Lee, C. J. Won, K. D. Lee, N. Hur, J. L. Chen, D. -Y. Cho, and T. W. Noh, Hybridized orbital states in spin-orbit coupled 3d--5d double perovskites studied by x-ray absorption spectroscopy, Phys. Rev. B {\bf 97}, 125123 (2018).
   
\bibitem{Marco_PRB_15} M. A. Laguna-Marco, P. Kayser, J. A. Alonso, M. J. Mart\'inez-Lope, M. V. Veenendaal, Y. Choi, and D. Haskel,  Electronic structure, local magnetism, and spin-orbit effect of Ir(IV)-, Ir(V)-, and Ir(VI)-based compounds, Phys. Rev. B {\bf 91}, 214433 (2015).

\bibitem{Vogl_PRB_18} M. Vogl, L. T. Corredor, T. Dey, R. Morrow, F. Scaravaggi, A. U. B. Wolter, S. Aswartham, S. Wurmehl, and B. B\"uchner, Interplay of 3d-and 5d-sublattice magnetism in the double perovskite substitution series La$_2$Zn$_{1-x}$Co$_x$IrO$_6$, Phys. Rev. B {\bf 97}, 035155 (2018).

\bibitem{Coutrim_PRB_16} L. T. Coutrim, E. M. Bittar, F. Stavale, F. Garcia, E. Baggio-Saitovitch, M. Abbate, R. J. O. Mossanek, H. P. Martins, D. Tobia, P. G. Pagliuso, and L. Bufai\c cal, Compensation temperatures and exchange bias in La$_{1.5}$Ca$_{0.5}$CoIrO$_6$, Phys. Rev. B {\bf 93}, 174406 (2016).
  
\bibitem{Kolchinskaya_PRB_12} A. Kolchinskaya, P. Komissinskiy, M. B. Yazdi, M. Vafaee, D. Mikhailova, N. Narayanan, H. Ehrenberg, F. Wilhelm, A. Rogalev, and L. Alff, Magnetism and spin-orbit coupling in Ir-based double perovskites La$_{2-x}$Sr$_x$CoIrO$_6$, Phys. Rev. B {\bf 85}, 224422 (2012).

\bibitem{Madhogaria_PRB_19} R. P. Madhogaria, R. Das, E. M. Clements, V. Kalappattil, M. H. Phan, H. Srikanth, N. T. Dang, D. P. Kozlenko, and N. S. Bingham, Evidence of long-range ferromagnetic order and spin frustration effects in the double perovskite La$_2$CoMnO$_6$, Phys. Rev. B {\bf 99}, 104436 (2019).

\bibitem{Coutrim_PRB_19} L. T. Coutrim, D. Rigitano, C. Macchiutti, T. J. A. Mori, R. Lora-Serrano, E. Granado, E. Sadrollahi, F. J. Litterst, M. B. Fontes, E. Baggio-Saitovitch, E. M. Bittar, and L. Bufai\c cal, Zero-field-cooled exchange bias effect in phase-segregated, La$_{2-x}A_x$CoMnO$_{6-\delta}$ ($A=$ Ba, Ca, Sr; $x=$ 0, 0.5), Phys. Rev. B {\bf 100}, 054428 (2019).

\bibitem{SahooPRB_19} R. C. Sahoo, Y. Takeuchi, A. Ohtomo, and Z. Hossain, Exchange bias and spin glass states driven by antisite disorder in the double perovskite compound LaSrCoFeO$_6$, Phys. Rev. B {\bf 100}, 214436 (2019).

\bibitem{Pradheesh_EPJB_12} R. Pradheesh, H. S. Nair, V. Sankaranarayanan, and K. Sethupathi, Large magnetoresistance and Jahn-Teller effect in Sr$_2$FeCoO$_6$, Eur. Phys. J. B {\bf 85}, 260 (2012).

\bibitem{MurthyJPD14} J. K. Murthy and A. Venimadhav, Multicaloric effect in multiferroic Y$_2$CoMnO$_6$, J. Phys. D: Appl. Phys. {\bf 47}, 445002 (2014).

\bibitem{SharmaAPL13} G. Sharma, J. Saha, S. D. Kaushik, V. Siruguri, and S. Patnaik, Magnetism driven ferroelectricity above liquid nitrogen temperature in Y$_2$CoMnO$_6$, Appl. Phys. Lett. {\bf 103}, 012903 (2013).

\bibitem{TanwarPRB19} K. Tanwar, D. S. Gyan, S. Bhattacharya, S. Vitta, A. Dwivedi, and T. Maiti, Enhancement of thermoelectric power factor by inducing octahedral ordering in La$_{2-x}$Sr$_x$CoFeO$_6$ double perovskites, Phys. Rev. B {\bf 99}, 174105 (2019).

\bibitem{ErtenPRL11} O. Erten, O. N. Meetei, A. Mukherjee, M. Randeria, N. Trivedi, and P. Woodward, Theory of half-metallic ferrimagnetism in double perovskites, Phys. Rev. Lett. {\bf 107}, 257201 (2011). 

\bibitem{Shafeie_JSSC_11} S. Shafeie, J. Grins, S. Y. Istomin, L. Karvonen, S. A. Chen, T. H. Chen, J. M. Chen, A. Weidenkaff, M. Karppinen, T. Sirtl, and G. Svensson, Phase formation, crystal structures and magnetic properties of perovskite-type phase in the system La$_2$Co$_{1+z}$(Mg$_x$Ti$_{1-x}$)$_{1-z}$O$_6$, J. Solid State Chem. {\bf 184}, 177 (2011).

\bibitem{Kobayashi_JPSJ_12} Y. Kobayashi, M. Kamogawa, Y. Terakado, and K. Asai, Magnetic properties of double perovskites  (Sr$_{1-x}$La$_{x}$)$_2$CoMO$_6$ with M = Sb, Nb, and Ta, J. Phys. Soc. Jpn. {\bf 81}, 044708 (2012).

 \bibitem{Yoshii_JSSC_2000} K. Yoshii, Magnetic transition in the perovskite Ba$_2$CoNbO$_6$, J. Solid State Chem. {\bf 151}, 294 (2000).
 
 \bibitem{ShuklaPRB18} R. Shukla and R. S. Dhaka, Anomalous magnetic and spin glass behavior in Nb substituted LaCo$_{1-x}$Nb$_x$O$_3$, Phys. Rev. B {\bf 97}, 024430 (2018).

\bibitem{RaviJALCOM18} R. Prakash, R. Shukla, P. Nehla, A. Dhaka, and R. S. Dhaka, Tuning ferromagnetism and spin state in La$_{(1-x)}$A$_x$CoO$_3$ (A =Sr, Ca) nanoparticles, J. Alloys Compd. {\bf 764}, 379 (2018).

\bibitem{ShuklaJPCC19} R. Shukla, A. Jain, M. Miryala, M. Murakami, K. Ueno,
S. M. Yusuf, and R. S. Dhaka, Spin dynamics and unconventional magnetism in insulating La$_{1-2x}$Sr$_{2x}$Co$_{1-x}$Nb$_x$O$_3$, J. Phys. Chem. C {\bf 123}, 22457 (2019).

\bibitem{Azcondo_Dalton_15} M. T. Azcondo, J. R. de Paz, K. Boulahya, C. Ritter, F. Garc\'ia-Alvarado, and U. Amador, Complex magnetic behavior of Sr$_2$CoNb$_{1-x}$Ti$_x$O$_6$ (0$\leqslant x\leqslant$0.5) as a result of a flexible microstructure, Dalton Trans. {\bf 44}, 3801 (2015).

\bibitem{Phan_JMMM_07} M. -H. Phan and S. C. Yu, Review of magnetocloric effect in manganite materials, J. Magn. Magn. Mater. {\bf 308}, 325 (2007).

\bibitem{Neumann_TA_06} A. Neumann and D. Walter, The thermal transformation from lanthanum hydroxide to lanthanum hydroxide oxide, Thermochim. Acta {\bf 445}, 200 (2006).

\bibitem{Carvajal_PB_93} J. Rodr\'iguez-Carvajal, Recent advances in magnetic structure determination by neutron powder diffraction, Physica B {\bf 192}, 55 (1993).

\bibitem{Glazer_AC_72} A. M. Glazer, The classification of tilted octahedra in perovskites, Acta Cryst. {\bf B28}, 3384 (1972).

\bibitem{Woodward_AC_97} P. M. Woodward, Octahedral tilting in perovskites. I. geometrical considerations, Acta Cryst. {\bf B53}, 32 (1997).

\bibitem{Shannon_AC_76} R. D. Shannon, Revised effective ionic radii and systematic studies of interatomic distances in halides and chalcogenides, Acta. Cryst. A {\bf 32}, 751 (1976). 

\bibitem{Barnes_AC_06} P. W. Barnes, M. W. Lufaso and P. M. Woodward, Structure determination of A$_2$M$^{3+}$TaO$_6$ and A$_2$M$^{3+}$NbO$_6$ ordered perovskites: octahedral tilting and pseudosymmetry, Acta Cryst. {\bf B62}, 384 (2006).

\bibitem{Andrews_Dalton_15} R. L. Andrews, A. M. Heyns, and P. M. Woodward, Raman studies of A$_2$MWO$_6$ tungstate double perovskites, Dalton Trans. {\bf 44}, 10700 (2015).

\bibitem{Iliev_PRB_07} M. N. Iliev, M. V. Abrashev, A. P. Litvinchuk, V. G. Hadjiev, H. Guo, and A. Gupta, Raman spectroscopy of ordered double perovskite La$_2$CoMnO$_6$ thin films, Phys. Rev. B {\bf 75}, 104118 (2007).

\bibitem{Prosandeev_PRB_05} S. A. Prosandeev, U. Waghmare, I. Levin, and J. Maslar, First-order Raman spectra of AB$^\prime_{1/2}$B$^{\prime\prime}_{1/2}$O$_3$ double perovskites, Phys. Rev. B {\bf 71}, 214307 (2005).

\bibitem{Kumaar_PRB_12} P. A. Kumar, R. Mathieu, R. Vijayaraghavan, S. Majumdar, O. Karis, P. Nordblad, B. Sanyal, O. Eriksson, and D. D. Sarma,  Ferrimagnetism, antiferromagnetism, and magnetic frustration in La$_{2-x}$Sr$_x$CuRuO$_6$ (0 $\leqslant$ x $\leqslant$ 1), Phys. Rev. B {\bf 86}, 094421 (2012).

\bibitem{Knizek_PRB_12} K. Kn\'i\v zek, J. Hejtm\'anek, M. Mary\v sko, Z. Jir\'ak, and J. Bur\v s\'ik, Stablization of the high spin state of Co$^{3+}$ in LaCo$_{1-x}$Rh$_x$O$_3$, Phys. Rev. B {\bf 85}, 134401 (2012).

\bibitem{MerzPRB10} M. Merz, P. Nagel, C. Pinta, A. Samartsev, H. v. L\"{o}hneysen, M. Wissinger, S. Uebe, A. Assmann, D. Fuchs, and S. Schuppler, Phys. Rev. B {\bf 82}, 174416 (2010).

\bibitem{BollettaPRB18} J. P. Bolletta, F. Pomiro, R. D. S\'anchez, V. Pomjakushin, G. Aurelio, A. Maignan, C. Martin, and R. E. Carbonio, Spin reorientation and metamagnetic transitions in RFe$_{0.5}$Cr$_{0.5}$O$_3$ perovskites (R = Tb, Dy,Ho, Er), Phys. Rev. B {\bf 98}, 134417 (2018).

\bibitem{Arrott_PR_57} A.  Arrott, Criterion for ferromagnetism from observations of magnetic isotherms, Phys. Rev. {\bf 108}, 1394 (1957).

\bibitem{Aharony_PRL_80} A. Aharony and E. Pytte, Infinite susceptibility phase in random uniaxial anisotropy magnets, Phys. Rev. Lett. {\bf 45}, 1583 (1980).

\bibitem{Banerjee_PL_64} S. K. Banerjee, On a generalised approach to first and second order magnetic transitions, Phys. Lett. {\bf 12}, 16 (1964).

\bibitem{Midya_PRB_11} A. Midya, S. N. Das, and P. Mandal, Anisotropic magnetic properties and giant magnetocaloric effect in antiferromagnetic RMnO$_3$ crystals (R =Dy, Tb, Ho, and Yb), Phys. Rev. B {\bf 84}, 235127 (2011).

\bibitem{Kumar_PRB_08} P. Kumar, N. K. Singh, K. G. Suresh, and A. K. Nigam, Magnetocaloric and magnetotransport properties of R$_2$Ni$_2$Sn compounds (R = Ce, Nd, Sm, Gd, and Tb), Phys. Rev. B {\bf 77}, 184411 (2008).

\bibitem{Emre_PRB_08} B. Emre, S. Aksoy, O. Posth, M. Acet, E. Duman, J. Lindner, and Y. Elerman, Antiferromagnetic-ferromagnetic crossover in La$_{0.5}$Pr$_{0.5}$Mn$_2$Si$_2$ and its consequences on magnetoelastic and magnetocaloric properties, Phys. Rev. B {\bf 78}, 144408 (2008).

\bibitem{Biswas_JAP_13} A. Biswas, S. Chandra, T. Samanta, M. H. Phan, I. Das, and H. Srikanth, The universal behavior of inverse magnetocaloric effect in antiferromagnetic materials, J. Appl. Phys. {\bf 113}, 17A902 (2013).

\bibitem{Krenke_NM_05} T. Krenke, E. duman, M. Acet, E. F. Wassermann, X. Moya, L. ma\~nosa, and A. Planes, Inverse magnetocaloric effect in ferromagnetic Ni-Mn-Sn alloys, Nature Mater {\bf 4}, 450 (2005).

\bibitem{Egolf_Conf_05} Proceedings of the first international conference on magnetic refrigeration at room temperature, edited by P. W. Egolf, (international institute of refrigeration Paris, France, 2005). 

\bibitem{Franco_JAP_06} V. Franco, J. S. Bl\'azquez, and A. Conde, Field dependence of the magnetocaloric effect in materials with a second order phase transition: A master curve for the magnetic entropy change, Appl. Phys. Lett. {\bf 89}, 222512 (2006).

\bibitem{Arrott_PRL_67} A. Arrott and J. E. Noakes, Approximate equation of state for nickel near its critical temperature, Phys. Rev. Lett. {\bf 19}, 786 (1967).

\bibitem{Law_nature_18} J. Y. Law, V. Franco, L. M. Moreno-Ram\'irez, A. Conde, D. Y. Karpenkov, I. Radulov, K. P. Skokov, and O. Gutfleisch, A quantitative criterion for determining the order of magnetic phase transitions using the magnetocaloric effect, Nat. Commun. {\bf 9}, 2680 (2018).

\bibitem{SarkarPRB08} P. Sarkar, P. Mandal, A. K. Bera, S. M. Yusuf, L. S. Sharath Chandra, and V. Ganesan, Field-induced first-order to second-order magnetic phase transition in Sm$_{0.52}$Sr$_{0.48}$MnO$_3$, Phys. Rev. B {\bf 78}, 012415 (2008).

\bibitem{XU_JCSJ_16} S. Xu, X. Lin, B. Ge, D. Ai, and Z. Peng, Conducting mechanism of Sr$_2$CoRO$_6$ (R = Mo, Nb) under different P$_{\rm O2}$, J. Ceram. Soc. Jpn. {\bf 124}, 813 (2016).

\bibitem{Paul_PRL_73} D. K. Paul and S. S. Mitra, Evaluation of Mott's parameters for hopping conduction in amorphous Ge, Si, and Se-Si, Phys. Rev. Lett. {\bf 31}, 1000 (1973).

\bibitem{Tong_PRB_04} W. Tong, B. Zhang, S. Tan, and Y. Zhang, Probability of double exchange between Mn and Fe in LaMn$_{1-x}$Fe$_x$O$_3$, Phys. Rev. B {\bf 70}, 014422 (2004).

\end{thebibliography}
\end{document}